\documentclass[%
 aip,
 amsmath,amssymb,
 reprint,%
]{revtex4-1}

\usepackage{graphicx}
\usepackage{dcolumn}
\usepackage{bm}

\usepackage[utf8]{inputenc}
\usepackage[T1]{fontenc}
\usepackage{mathptmx}
\usepackage{etoolbox}

\usepackage{listings}

\usepackage{xcolor}

\usepackage{graphicx}

\definecolor{vscodegreen}{HTML}{6A9955}

\lstdefinestyle{mystyle}{
    commentstyle=\color{vscodegreen},
    keywordstyle=\color{blue},
    stringstyle=\color{purple},
    basicstyle=\ttfamily\footnotesize,
    breakatwhitespace=false,         
    breaklines=true,                 
    captionpos=b,                    
    keepspaces=true,                 
    numbersep=5pt,                  
    showspaces=false,                
    showstringspaces=false,
    showtabs=false,                  
    tabsize=2
}

\usepackage{float}
\usepackage{makecell}
\usepackage{hyperref}

\makeatletter
\def\@email#1#2{%
 \endgroup
 \patchcmd{\titleblock@produce}
  {\frontmatter@RRAPformat}
  {\frontmatter@RRAPformat{\produce@RRAP{*#1\href{mailto:#2}{#2}}}\frontmatter@RRAPformat}
  {}{}
}%
\makeatother
\begin{document}

\preprint{AIP/123-QED}

\title[An efficient hp-Variational PINNs framework for incompressible Navier-Stokes equations]{An efficient hp-Variational PINNs framework for incompressible Navier-Stokes equations}
\author{Thivin Anandh}
 \affiliation{Department of Computational and Data Sciences, Indian Institute of Science, Bangalore}

\author{Divij Ghose}
\affiliation{Department of Computational and Data Sciences, Indian Institute of Science, Bangalore}

\author{Ankit Tyagi}
\affiliation{
Shell India Markets Private Limited%
}%
\author{Abhineet Gupta}
\affiliation{
Shell India Markets Private Limited%
}%

\author{Suranjan Sarkar}
\affiliation{
Shell India Markets Private Limited%
}%

\author{Sashikumaar Ganesan*}
 \email{sashi@iisc.ac.in}
\affiliation{%
Department of Computational and Data Sciences, Indian Institute of Science, Bangalore%
}%

\date{\today}

\begin{abstract}
Physics-informed neural networks (PINNs) are able to solve partial differential equations (PDEs) by incorporating the residuals of the PDEs into their loss functions. Variational Physics-Informed Neural Networks (VPINNs) and hp-VPINNs use the variational form of the PDE residuals in their loss function. Although hp-VPINNs have shown promise over traditional PINNs, they suffer from higher training times and lack a framework capable of handling complex geometries, which limits their application to more complex PDEs. As such, hp-VPINNs have not been applied in solving the Navier-Stokes equations, amongst other problems in CFD, thus far. FastVPINNs was introduced to address these challenges by incorporating tensor-based loss computations, significantly improving the training efficiency. Moreover, by using the bilinear transformation, the FastVPINNs framework was able to solve PDEs on complex geometries. In the present work, we extend the FastVPINNs framework to vector-valued problems, with a particular focus on solving the incompressible Navier-Stokes equations for two-dimensional forward and inverse problems, including problems such as the lid-driven cavity flow, the Kovasznay flow, and flow past a backward-facing step for Reynolds numbers up to 200. Our results demonstrate a  2x improvement in training time while maintaining the same order of accuracy compared to PINNs algorithms documented in the literature. We further showcase the framework's efficiency in solving inverse problems for the incompressible Navier-Stokes equations by accurately identifying the Reynolds number of the underlying flow. Additionally, the framework's ability to handle complex geometries highlights its potential for broader applications in computational fluid dynamics. This implementation opens new avenues for research on hp-VPINNs, potentially extending their applicability to more complex problems.

\end{abstract}

\maketitle


\section{Introduction}
\label{sec:introduction}

Traditional low-fidelity models are fast but ignore physical effects, leading to significant uncertainties in project design and process optimization. First-principle based numerical simulations capture all physics effects, providing more accurate predictions and reducing uncertainties. While high-fidelity simulations greatly improve prediction accuracy, they are computationally expensive. Machine learning offers a promising alternative, but often requires huge amount of data that may not be available for scientific and engineering problems. In recent years, physics-informed neural networks (PINNs) have emerged as a powerful tool for solving partial differential equations (PDEs) in various scientific domains. Since their introduction by Raissi et al.\cite{raissi2019physics}, PINNs have found applications in fields ranging from climate modeling \cite{de2021assessing} to solid mechanics \cite{haghighat2021physics}. Karniadakis et al.~\cite{karniadakis2021physics} provide a comprehensive review on physics-informed machine learning, discussing its capabilities in integrating noisy data with mathematical models, addressing high-dimensional problems, and discovering hidden physics in both forward and inverse problems. 

The field of fluid dynamics, in particular, has seen significant advancements through the use of PINNs \cite{cai2021physics,arthurs2021active,raissi2020hidden, mahmoudabadbozchelou2022nn, biswas2023three}. A notable development in this area is the NSFNets framework \cite{jin2021nsfnets}, which focused on solving incompressible Navier-Stokes equations using different formulations such as Velocity-Pressure (VP) and Vorticity-Velocity (VV), while also investigating the impact of loss function weighting. Further studies have extended the application of PINNs to 3D laminar flows~\cite{biswas2023three} and turbulent flows using Reynolds Averaged Navier-Stokes~(RANS) equations \cite{eivazi2022physics}. Moreover, the use of PINNs have been studied for fluid applications, such as novel training methodologies for turbulent flows~\cite{kag2022physics}, dynamic weighting strategies~\cite{li2022dynamic}, multiphase modeling with heat transfer~\cite{zhang2024physics}, and compressible flow equations~\cite{wassing2024physics}. More recently, domain decomposition strategies for training Navier-Stokes equations with PINNs have been investigated using interface loss conditions \cite{gu2024physics}.

Variational Physics-Informed Neural Networks (VPINNs), which use the variational form of the PDE in the loss function, have shown promise in solving PDEs~\cite{kharazmi2019variational, khodayi2020varnet}. Moreover, concepts like h-refinement and p-refinement can be applied to VPINNs to further increase accuracy, resulting in the hp-VPINNs framework\cite{kharazmi2021hp}. However, despite their benefits, hp-VPINNs have not been succesfully applied in CFD owing to several limitations. Firstly, training hp-VPINNs is computationally expensive due to the non-optimal implementation of the residual calculation. This means that computation times become unrealistic as the number of cells increases. Secondly, current frameworks are limited to uniform meshes and are unable to perform computations on complex, irregular meshes that are often seen in CFD applications. FastVPINNs proposed by Anandh et al.~\cite{anandh2024fastvpinns, anandh2024fastvpinnsjoss, ghose2024fastvpinns} addressed the challenges in existing implementation and provided a 100x speedup in training time compared to the existing implementation of hp-VPINNs~\cite{hp_vpinns_github}. In this work, we extend the implementation of FastVPINNs for CFD applications. We demonstrate the ability of our framework by solving the 2D incompressible Navier-Stokes equation, an application not previously addressed using hp-VPINNs in the literature. We present results for the solution of various flow scenarios, including Kovasznay flow, lid-driven cavity, and Falkner-Skan problem. We analyze the accuracy of our solutions by comparing them with existing literature or finite element method (FEM) results. Additionally, we also compare the training times of FastVPINNs with other PINNs frameworks mentioned in the literature for solving incompressible Navier-Stokes equations. 


\section{Preliminaries}
\label{sec:preliminaries}
In this section, we will briefly explain the FastVPINNs methodology for solving scalar PDEs using hp-VPINNs and then state the tensor-based loss computation routines used in FastVPINNs.

\subsection{Variational form of a scalar PDE}
We begin with the Poisson problem in 2D with homogeneous Dirichlet boundary conditions, defined on the open and bounded domain $\Omega \subset \mathbb{R}^2$
\begin{align}
    \begin{split}
        -\Delta u &= f \quad \text{in} \ \Omega, \\
        u &= g \quad \text{on} \ \partial \Omega,
    \end{split}
    \label{eq:poisson2d}
\end{align}
where $u$ is an unknown scalar solution, $f$ is a known and sufficiently smooth source function, and $g$ is the Dirichlet boundary condition imposed on the domain boundary, $\partial \Omega$. Let $\text{H}^1(\Omega)$ denote the conventional Sobolev space. The variational form of Eq.~\eqref{eq:poisson2d} can be formulated as:\\

\noindent Find $u \in V$ such that,
\[
a(u,v) = f(v) \quad \text {for all } v \in V, 
\]
where,
\[
 V:=\left\{v \in \text{\text{H}}^1(\Omega) : v = 0 \quad \text{on} \quad \partial \Omega \right\},
\]
and 
\begin{equation}
    a(u,v) := \int_{\Omega} \nabla u \cdot \nabla v \; dx,  \qquad
        f(v) := \int_{\Omega} fv \; dx.
    \label{eqn:Poisson2D_weakform}
\end{equation}
The domain $\Omega$ is then divided into non-overlapping cells, labeled as $K_k$, where $k=1,2,\ldots,\texttt{N\_{elem}}$, ensuring that the complete union $\bigcup_{k=1}^\texttt{N\_{elem}}K_k = \Omega$. We define $V_h$ as a finite-dimensional subspace of $V$, spanned by the basis functions $\phi_h := \{\phi_j(x)\},\; j = 1,2,\ldots,\texttt{N\_{test}}$, where $\texttt{N\_{test}}$ indicates the dimension of $V_h$. As a result, the discretized variational formulation related to Eq.~\eqref{eqn:Poisson2D_weakform} can be written as follows,\\

\noindent Find $u_h \in V_h$ such that,
\begin{equation}
a_h(u_h,v) = f_h(v) \quad \text {for all } v \in V_h,\label{eqn:Poisson2D_disform}
\end{equation}
where,
\[
    a_h(u_h,v) := \sum_{k=1}^{\texttt{N\_{elem}}}\int_{K_k} \nabla u_h \cdot \nabla v \; dK,  \qquad
        f_h(v) := \sum_{k=1}^{\texttt{N\_{elem}}}\int_{K_k} fv \; dK.
\]
These integrals are approximated using numerical quadrature as 
\begin{align*}
    \int_{K_k} \nabla u_h \cdot \nabla v \; dK &\approx  \sum_{q=1}^{\texttt{N\_quad}} w_q ~\nabla u_h(x_q) \cdot \nabla v(x_q)\;,  \\
       \int_{K_k} fv \; dK & \approx   \sum_{q=1}^{\texttt{N\_quad}}w_q ~f(x_q)\,v(x_q)\;.
\end{align*}
Here, \texttt{N\_quad} is the number of quadrature points in a element. $x_q$ and $w_q$ are the coordinates and the weights of the quadrature point q respectively.

\subsection{hp-Variational Physics Informed Neural Networks}\label{sec:hp-VPINNs}
In hp-Variational Physics Informed Neural Networks (hp-VPINNs) \cite{kharazmi2021hp}, $u_h$ is approximated by a neural network, 
\begin{equation*}
    u_h(x) \approx u_{\text{NN}}(\mathbf{x}; W,b), 
\end{equation*}
where $W$ and $b$ represent the weights and biases of the network. By substituting $u_{\text{NN}}$ in Eq.~\eqref{eqn:Poisson2D_disform}, we can obtain the element-wise residuals, 
\begin{equation*}
    \begin{split}
        \mathcal{W}_k(\mathbf{x};W,b)&= \int_{K_k} \left( \nabla u_{\text{NN}}(\mathbf{x};W,b) \cdot \nabla v_k ~ - ~ f\,v_k\right)\,dK,
    \end{split}
\label{eqn:VPINNs_residual}
\end{equation*}
which can then be summed over all elements to give the variational loss, 
\begin{equation} \label{eq:variational_loss}
    L_v(W,b) = \frac{1}{\texttt{\scriptsize N\_elem}}\sum_{k=1}^{\texttt{N\_elem}}\left|\mathcal{W}_k(\mathbf{x};W,b)\right|^2
\end{equation}
The neural network is trained using the cost function
\begin{equation*}
    L_{\text{VPINN}}(W,b) = L_v + \tau L_b
    \label{eqn:vpinns_final_loss_fn_abbr}
\end{equation*}
where, $L_b$ is the Dirichlet boundary loss, 
\begin{align*}
    \begin{split}
        L_b(W,b) &= \frac{1}{N_D}\sum_{d=1}^{N_{D}}\left(u_{\text{NN}}(\mathbf{x}; W, b) - g(x)\right)^2. \quad \text{on} \ \partial \Omega.
    \end{split}
    \label{eqn:PINNs_Loss_components}
\end{align*}
Here, $\tau$ is a scaling factor which controls the penalty on the boundary loss and $N_D$ is the number of Dirichlet boundary points.

\subsection{The FastVPINNs framework}\label{sec:fastvpinns}
\begin{figure}[!htp]
    \centering
    \includegraphics[width=0.45\textwidth]{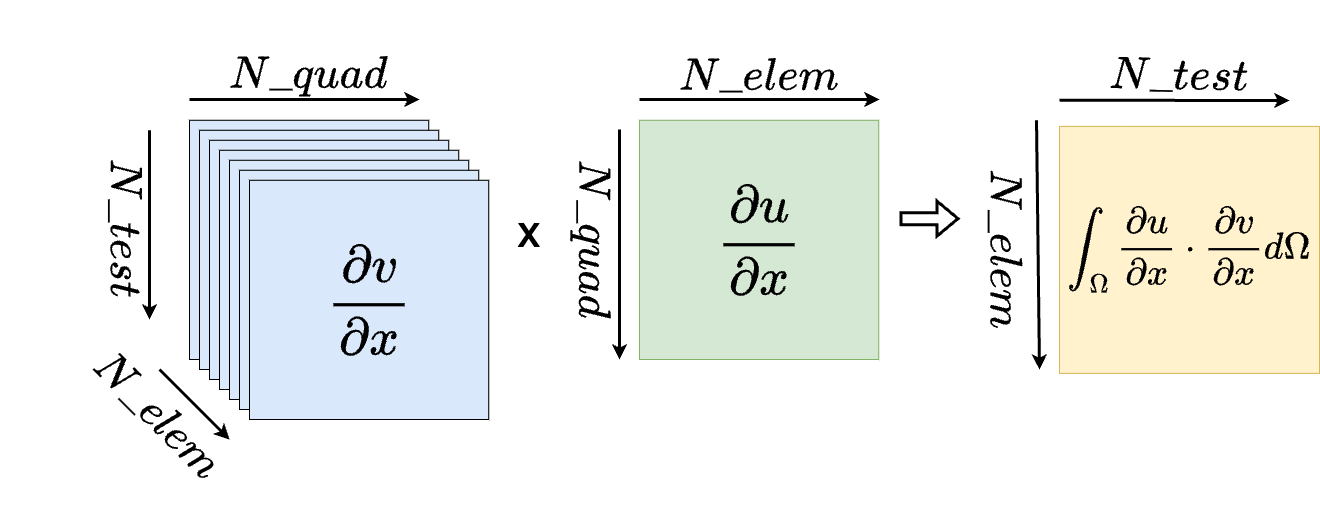}
    \caption{FastVPINNs Tensor schematic representation for residual computation.}
    \label{fig:fast_vpinns}
\end{figure}

The FastVPINNs framework~\cite{anandh2024fastvpinns, anandh2024fastvpinnsjoss, ghose2024fastvpinns} uses a tensor-based approach to calculate the variational loss to reduce the training time of hp-VPINNs. To solve the Poisson problem using FastVPINNs, we first pre-compute the gradient of the test functions in the actual domain and assemble them into the tensors $\boldsymbol{\mathcal{G}^x}, \boldsymbol{\mathcal{G}^y} \in \mathbb{R}^{\texttt{N\_elem} \times \texttt{N\_test} \times \texttt{N\_quad}}$,
\begin{align*}
    \begin{split}
        \boldsymbol{\mathcal{G}}^{\mathbf{x}}_{ijk} = \frac{\partial v_j}{\partial x}\bigg\rvert_{(x_{ik}, y_{ik})}, 
        \boldsymbol{\mathcal{G}}^{\mathbf{y}}_{ijk} = \frac{\partial v_j}{\partial y}\bigg\rvert_{(x_{ik}, y_{ik})}.
    \end{split}
\end{align*}
Here, $i = 1, 2, \cdots, \texttt{N\_elem}$, $j = 1, 2, \cdots, \texttt{N\_test}$, and $k = 1, 2, \cdots, \texttt{N\_quad}$. The test functions are computed at the $k^{th}$ quadrature point of the $i^{th}$ element. These test functions are computed on the reference element and transferred to the corresponding actual element using bilinear transformation. Figure~\ref{fig:fast_vpinns} shows the tensor-based loss computation schematic of FastVPINNs.

The gradients of the output of the neural network are arranged in the matrices $\mathbf{u^x}, \mathbf{u^y} \in \mathbb{R}^{\texttt{N\_elem} \times\texttt{N\_quad}}$, 
\begin{equation*}
        \mathbf{u}^{\mathbf{x}}_{ik} = \frac{\partial u_{\text{NN}}}{\partial x}\bigg\rvert_{(x_{ik}, y_{ik})}, 
        \mathbf{u}^{\mathbf{y}}_{ik} = \frac{\partial u_{\text{NN}}}{\partial y}\bigg\rvert_{(x_{ik}, y_{ik})},
\end{equation*}
and the values of the source term, $f$, at each quadrature point of each element is pre-computed and stored in the matrix $\mathbf{F} \in \mathbb{R}^{\texttt{N\_elem} \times \texttt{N\_test}}$,
\begin{equation*}
    \mathbf{F}_{ij} = f(x_{ij}).
\end{equation*}
Finally, the variational loss in \eqref{eq:variational_loss} can be calculated
\begin{equation*}
\begin{aligned}
\mathbf{R}^{\mathbf{x}}_{ij} &= \sum_{k=1}^{\texttt{\scriptsize N\_quad}} \boldsymbol{\mathcal{G}}^{\mathbf{x}}_{ijk} \mathbf{u}^{\mathbf{x}}_{ik}, \\
\mathbf{R}^{\mathbf{y}}_{ij} &= \sum_{k=1}^{\texttt{\scriptsize N\_quad}} \boldsymbol{\mathcal{G}}^{\mathbf{y}}_{ijk} \mathbf{u}^{\mathbf{y}}_{ik}, \\
\mathbf{R}_{ij} &= \mathbf{R}^{\mathbf{x}}_{ij} + \mathbf{R}^{\mathbf{y}}_{ij} - \mathbf{F}_{ij}, \\
L_v &= \frac{1}{\texttt{\scriptsize N\_test}} \sum_{i=1}^{\texttt{\scriptsize N\_elem}} \sum_{j=1}^{\texttt{\scriptsize N\_test}} (\mathbf{R}_{ij})^2,
\end{aligned}
\end{equation*}
The operation required to calculate $\mathbf{R^x}$ and $\mathbf{R^y} \in \mathbb{R}^{\texttt{N\_elem} \times \texttt{N\_test}}$ is equivalent to a tensor contraction along the third dimension of quadrature points. This operation is efficiently performed on GPUs using the \texttt{tf.matvec} function provided by the Tensorflow-v2.0 library~\cite{tensorflow2015-whitepaper}. The pseudocode for this procedure can be found in the FastVPINNs paper~\cite{anandh2024fastvpinns}.

\section{Methodology}\label{sec:methodology}
We will first examine the incompressible Navier-Stokes equations, along with their weak forms, which are used in the loss computations. Following this, we will explore the extension of the existing FastVPINNs algorithm to vector-valued problems for incompressible Navier-Stokes equations.

\subsection{Incompressible Navier-Stokes Equation}
Consider the 2D steady state incompressible Navier-Stokes problem in a bounded domain $\Omega \subset \mathbb{R}^2$,
\begin{align*}
    \begin{split}
        -\frac{1}{Re} \Delta \mathbf{u} + (\mathbf{u}\cdot\nabla)\mathbf{u} + \nabla p &= \mathbf{f} \quad \text{in} \ \Omega , \\
        \nabla  \cdot \mathbf{u} &= 0  \quad \text{in} \ \Omega , \\
        \mathbf{u} &= \mathbf{g} \quad \text{on} \ \partial \Omega_D, \\
        \nabla \mathbf{u} \cdot \mathbf{n} &= \mathbf{h} \quad \text{on} \ \partial \Omega_N,
    \end{split}
\end{align*}
where $\mathbf{u} = (u, v)$ denotes the fluid velocity, $\mathbf{f} = (f_x, f_y)$ denotes the forcing terms, $p$ represents the pressure, $Re$ is the Reynolds number where $Re = \frac{1}{\nu}$, and $\nu$ is the kinematic viscosity. The Dirichlet boundary condition is imposed on the domain boundary $\partial \Omega_D$, where the prescribed velocity field is given by $\mathbf{g} = (g_x, g_y)$. The Neumann boundary condition is imposed on $\partial \Omega_N$, where $\mathbf{h} = (h_x, h_y)$ specifies the normal derivative of velocity.

The weak formulation of the incompressible Navier-Stokes equations can be derived by multiplying each equation with appropriate test functions and integrating over the domain. For the momentum equations, we use a vector-valued test function $\mathbf{\phi} = (\phi_x, \phi_y)$, while for the continuity equation, we use a scalar test function $\psi$. This approach allows us to transform the strong form of the equations into a variational form. By applying integration by parts to the viscous and pressure terms, we obtain the following weak formulation:
\begin{align*}
     \frac{1}{Re} \int_{\Omega} \left(\nabla \mathbf{u} : \nabla \mathbf{\phi} \right) \, d\Omega 
    + \int_{\Omega} ((\mathbf{u} \cdot \nabla)\mathbf{u}) \cdot \mathbf{\phi} \, d\Omega \nonumber \\
    - \int_{\Omega} p (\nabla \cdot \mathbf{\phi}) \, d\Omega 
    &= \int_{\Omega} \mathbf{f} \cdot \mathbf{\phi} \, d\Omega, \\
    \int_{\Omega} (\nabla \cdot \mathbf{u}) \psi \, d\Omega &= 0.
\end{align*}
To better illustrate the component-wise nature of the weak formulation, we can expand the vector-valued notation into separate equations for each velocity component. This expansion results in three equations:

\begin{widetext}
\begin{align}
    \frac{1}{Re} \int_{\Omega}  \left(\frac{\partial u}{\partial x}\frac{\partial \phi_x}{\partial x} + \frac{\partial u}{\partial y}\frac{\partial \phi_x}{\partial y}\right) \, d\Omega 
    + \int_{\Omega} \left(u\frac{\partial u}{\partial x} + v\frac{\partial u}{\partial y}\right) \phi_x \, d\Omega 
    - \int_{\Omega} p \frac{\partial \phi_x}{\partial x} \, d\Omega 
    &= \int_{\Omega} f_x \phi_x \, d\Omega, \label{eq:weak_u} \\
    \frac{1}{Re} \int_{\Omega}  \left(\frac{\partial v}{\partial x}\frac{\partial \phi_y}{\partial x} + \frac{\partial v}{\partial y}\frac{\partial \phi_y}{\partial y}\right) \, d\Omega 
    + \int_{\Omega} \left(u\frac{\partial v}{\partial x} + v\frac{\partial v}{\partial y}\right) \phi_y \, d\Omega 
    - \int_{\Omega} p \frac{\partial \phi_y}{\partial y} \, d\Omega 
    &= \int_{\Omega} f_y \phi_y \, d\Omega, \label{eq:weak_v} \\
    \int_{\Omega} \left(\frac{\partial u}{\partial x} + \frac{\partial v}{\partial y}\right) \psi \, d\Omega &= 0. \label{eq:weak_p}
\end{align}
\end{widetext}
Here, Eqs.~\eqref{eq:weak_u}, \eqref{eq:weak_v}, and \eqref{eq:weak_p} represent the weak form of the momentum equation in the x-direction, the momentum equation in the y-direction, and the continuity equation for 2D incompressible flow, respectively. It should be noted that in typical finite element formulations, the orders of the test functions for velocity and pressure are chosen to satisfy the inf-sup condition, also known as the Ladyzhenskaya-Babuška-Brezzi (LBB) condition~\cite{babuvska1973finite, brezzi1974existence}. This selection ensures stability of the numerical solution for flow problems. However, for the purposes of this paper, we have employed equal-order finite elements for both velocity and pressure. This simplification allows us to focus on the core aspects of the method while acknowledging that it may not provide optimal stability for all flow regimes. For more information on the inf-sup condition for the Navier-Stokes equation, refer to Ganesan et al~\cite{ganesan2017finite}.

\subsection{Loss Formulation of hp-VPINNs for Incompressible Navier-Stokes Equation}

In hp-VPINNs, the solution vector reads
\begin{equation*}
    \begin{pmatrix}
        u\\
        v \\
        p
        
    \end{pmatrix} \approx \mathbf{u}_{\text{NN}}(\mathbf{x}; W,b) = 
    \begin{pmatrix}
        u_{\text{NN}}(\mathbf{x}; W,b) \\
        v_{\text{NN}}(\mathbf{x}; W,b) \\
        p_{\text{NN}}(\mathbf{x}; W,b)
    \end{pmatrix},
\end{equation*}
where $W$ and $b$ represent the weights and biases of the network. By substituting the components of $\mathbf{u}_{\text{NN}}$ into Eqs.~\eqref{eq:weak_u}, \eqref{eq:weak_v}, and \eqref{eq:weak_p}, we can obtain the element-wise residuals for each equation of the Navier-Stokes system as:

\begin{widetext}
\begin{align*}
    \mathcal{W}_k^u(\mathbf{x};W,b) &= \int_{K_k} \left[\frac{1}{Re} \left(\frac{\partial u_{\text{NN}}}{\partial x}\frac{\partial \phi_x}{\partial x} + \frac{\partial u_{\text{NN}}}{\partial y}\frac{\partial \phi_x}{\partial y}\right) + \left(u_{\text{NN}}\frac{\partial u_{\text{NN}}}{\partial x} + v_{\text{NN}}\frac{\partial u_{\text{NN}}}{\partial y}\right) \phi_x - p_{\text{NN}} \frac{\partial \phi_x}{\partial x}\right] \,dK - \int_{K_k} f_x \phi_x \,dK,
     \\
    \mathcal{W}_k^v(\mathbf{x};W,b) &= \int_{K_k} \left[\frac{1}{Re} \left(\frac{\partial v_{\text{NN}}}{\partial x}\frac{\partial \phi_y}{\partial x} + \frac{\partial v_{\text{NN}}}{\partial y}\frac{\partial \phi_y}{\partial y}\right) + \left(u_{\text{NN}}\frac{\partial v_{\text{NN}}}{\partial x} + v_{\text{NN}}\frac{\partial v_{\text{NN}}}{\partial y}\right) \phi_y - p_{\text{NN}} \frac{\partial \phi_y}{\partial y}\right] \,dK - \int_{K_k} f_y \phi_y \,dK,
    \\
    \mathcal{W}_k^p(\mathbf{x};W,b) &= \int_{K_k} \left(\frac{\partial u_{\text{NN}}}{\partial x} + \frac{\partial v_{\text{NN}}}{\partial y}\right) \psi \,dK.
\end{align*}
\end{widetext}

These terms are summed over all elements to obtain the variational loss for the Navier-Stokes equation which reads, 

\begin{align*}
    L^u(W,b) &= \frac{1}{\texttt{\scriptsize N\_elem}}\sum_{k=1}^{\texttt{N\_elem}}|\mathcal{W}_k^u(\mathbf{x};W,b)|^2, \\
    L^v(W,b) &= \frac{1}{\texttt{\scriptsize N\_elem}}\sum_{k=1}^{\texttt{N\_elem}}|\mathcal{W}_k^v(\mathbf{x};W,b)|^2,  \\
    L^c(W,b) &= \frac{1}{\texttt{\scriptsize N\_elem}}\sum_{k=1}^{\texttt{N\_elem}}|\mathcal{W}_k^p(\mathbf{x};W,b)|^2, 
\end{align*}

\begin{equation} \label{eq:variational_loss_NSE}
    L_v^{\text{NSE}}(W,b) = \alpha L^u(W,b) + \beta L^v(W,b) + \gamma L^c(W,b).
\end{equation}
Here, $L^u(W,b)$ and $L^v(W,b)$ are the variational losses for the x- and y- momentum components, respectively, while $L^c(W,b)$ is the variational loss for the continuity equation. The terms $\alpha$, $\beta$, and $\gamma$ are weights added to the loss components of the Navier-Stokes equations. These weights allow for fine-tuning the relative importance of each physical constraint in the overall loss function, potentially improving the convergence of the neural networks.
The Dirichlet boundary loss $L_d$ for each component of velocity in the Navier-Stokes equations is represented as:
\begin{align*}
    L_d^u(W,b) &= \frac{1}{N_D}\sum_{d=1}^{N_{D}}\left(u_{\text{NN}}(\mathbf{x}; W, b) - g_x(\mathbf{x})\right)^2,   \\
    L_d^v(W,b) &= \frac{1}{N_D}\sum_{d=1}^{N_{D}}\left(v_{\text{NN}}(\mathbf{x}; W, b) - g_y(\mathbf{x})\right)^2.
\end{align*}
The total Dirichlet boundary loss is then given by:
\begin{equation*}
    L_d^{\text{NSE}}(W,b) = L_d^u(W,b) + L_d^v(W,b) \label{eqn:NSE_Loss_Dirichlet_total}
\end{equation*}
The Neumann boundary loss $L_n$ for each component of velocity in the Navier-Stokes equations is represented as:
\begin{align*}
    L_n^u(W,b) &= \frac{1}{N_N}\sum_{n=1}^{N_{N}}\left(\frac{\partial u_{\text{NN}}}{\partial n}(\mathbf{x}; W, b) - h_x(\mathbf{x})\right)^2, \\
    L_n^v(W,b) &= \frac{1}{N_N}\sum_{n=1}^{N_{N}}\left(\frac{\partial v_{\text{NN}}}{\partial n}(\mathbf{x}; W, b) - h_y(\mathbf{x})\right)^2. 
\end{align*}
Here, $N_N$ represents the number of Neumann boundary points. The total Neumann boundary loss is then given by:
\begin{equation}
    L_n^{\text{NSE}}(W,b) = L_n^u(W,b) + L_n^v(W,b) \label{eqn:NSE_Loss_Neumann_total}
\end{equation}
The neural network is trained to minimize the total loss function, which reads:

\begin{equation*}
    L^{\text{NSE}}_{\text{VPINN}}(W,b) = L_v^{\text{NSE}} + \tau_D L_d^{\text{NSE}} + \tau_N L_n^{\text{NSE}}
    \label{eqn:NSE_vpinns_final_loss_fn}
\end{equation*}
Here, $L^{\text{NSE}}_{\text{VPINN}}(W,b)$ represents the total loss function for the Navier-Stokes equations. $L_v^{\text{NSE}}$ is the variational loss, $L_d$ is the Dirichlet boundary loss, and $L_n$ is the Neumann boundary loss. The terms $\tau_D$ and $\tau_N$ are weighting factors for the Dirichlet and Neumann boundary losses, respectively, allowing for balance between the enforcement of the physics in the domain and the satisfaction of boundary conditions.

It should be noted that the variational form of the PDE naturally gives rise to Neumann boundary terms when performing integration by parts, a technique commonly used for imposing Neumann boundary conditions in solvers like FEM. However, in this FastVPINNs framework, we impose the Neumann boundary conditions on the Navier-Stokes equations as shown in Eq.~\eqref{eqn:NSE_Loss_Neumann_total}, which is similar to how PINNs enforce such conditions. This approach allows for a more direct and flexible implementation of Neumann conditions within the neural network-based solver.

\section{FastVPINNs Implementation for Incompressible Navier-Stokes Problem}\label{sec:implementation}
The loss computation for the Navier-Stokes equations follows a similar approach to that of scalar equations, as described in Section~\ref{sec:fastvpinns}. However, it is extended to accommodate the vector nature of the problem. The process involves assembling test functions and their gradients into tensors, while the predicted solution of the neural network and its gradients are organized into matrices. A tensor-matrix operation is then applied to generate residual matrices for each component of the Navier-Stokes equations: momentum in the x-direction, momentum in the y-direction, and the divergence equation. These residuals are subsequently reduced to scalar values and combined to form the total variational loss. This loss is then augmented with boundary losses to complete the overall loss function for the neural network. 
\begin{equation*}
\begin{aligned}
\mathbf{R}^{\text{diff}_x}_{ij} &= \frac{1}{\text{Re}} \sum_{k=1}^{\texttt{\scriptsize N\_quad}} (\boldsymbol{\mathcal{G}}^x_{ijk} \mathbf{u}^x_{ik} + \boldsymbol{\mathcal{G}}^y_{ijk} \mathbf{u}^y_{ik}), \\
\mathbf{R}^{\text{conv}_x}_{ij} &= \sum_{k=1}^{\texttt{\scriptsize N\_quad}} \boldsymbol{\mathcal{T}}_{ijk} (\mathbf{u}^x_{ik} \mathbf{u}_{ik} + \mathbf{u}^y_{ik} \mathbf{v}_{ik}), \\
\mathbf{R}^{\text{press}_x}_{ij} &= \sum_{k=1}^{\texttt{\scriptsize N\_quad}} \boldsymbol{\mathcal{G}}^x_{ijk} \mathbf{p}_{ik}, \\
\mathbf{R}^x_{ij} &= \mathbf{R}^{\text{diff}_x}_{ij} + \mathbf{R}^{\text{conv}_x}_{ij} - \mathbf{R}^{\text{press}_x}_{ij}.
\end{aligned}
\label{eqn:tensor_nse_loss_m_x}
\end{equation*}
The above equations summarise the loss computation of x-momentum component, where $\boldsymbol{\mathcal{G}}^x_{ijk}$ and $\boldsymbol{\mathcal{G}}^y_{ijk}$ are the test function gradient tensors in the $x$ and $y$ directions, $\boldsymbol{\mathcal{T}}_{ijk}$ is the test function tensor. The subscripts $i$, $j$, and $k$ represent the element, test function, and quadrature point indices, respectively. $\mathbf{u}_{ik}$, $\mathbf{v}_{ik}$, and $\mathbf{p}_{ik}$ are the predicted velocity components and pressure from the neural network, while $\mathbf{u}^x_{ik}$ and $\mathbf{u}^y_{ik}$ are x and y gradients of the velocity component $u_{NN}$, respectively. The residuals $\mathbf{R}^{\text{diff}_x}_{ij}$, $\mathbf{R}^{\text{conv}_x}_{ij}$, and $\mathbf{R}^{\text{press}_x}_{ij}$ represent the diffusion, convection, and pressure terms for the x-momentum equation.
\begin{equation*}
\begin{aligned}
\mathbf{R}^{\text{diff}_y}_{ij} &= \frac{1}{\text{Re}} \sum_{k=1}^{\texttt{\scriptsize N\_quad}} (\boldsymbol{\mathcal{G}}^x_{ijk} \mathbf{v}^x_{ik} + \boldsymbol{\mathcal{G}}^y_{ijk} \mathbf{v}^y_{ik}), \\
\mathbf{R}^{\text{conv}_y}_{ij} &= \sum_{k=1}^{\texttt{\scriptsize N\_quad}} \boldsymbol{\mathcal{T}}_{ijk} (\mathbf{v}^x_{ik} \mathbf{u}_{ik} + \mathbf{v}^y_{ik} \mathbf{v}_{ik}), \\
\mathbf{R}^{\text{press}_y}_{ij} &= \sum_{k=1}^{\texttt{\scriptsize N\_quad}} \boldsymbol{\mathcal{G}}^y_{ijk} \mathbf{p}_{ik}, \\
\mathbf{R}^y_{ij} &= \mathbf{R}^{\text{diff}_y}_{ij} + \mathbf{R}^{\text{conv}_y}_{ij} - \mathbf{R}^{\text{press}_y}_{ij}.
\end{aligned}
\end{equation*}
These equations represent the y-momentum component, analogous to the x-momentum component. $\mathbf{v}^x_{ik}$ and $\mathbf{v}^y_{ik}$ are the gradients of the y-component of velocity~($v_{NN}$).
\begin{equation*}
\begin{aligned}
\mathbf{R}^{\text{div}}_{ij} &= \sum_{k=1}^{\texttt{\scriptsize N\_quad}} (\boldsymbol{\mathcal{P}}^x_{ijk} \mathbf{u}^x_{ik} + \boldsymbol{\mathcal{P}}^y_{ijk} \mathbf{v}^y_{ik}),
\end{aligned}
\end{equation*}
This equation represents the divergence residual, where $\boldsymbol{\mathcal{P}}^x_{ijk}$ and $\boldsymbol{\mathcal{P}}^y_{ijk}$ are the pressure test function gradient tensors.
\begin{equation*}
\begin{aligned}
L^u &= \frac{1}{\texttt{\scriptsize N\_test}} \sum_{i=1}^{\texttt{\scriptsize N\_elem}} \sum_{j=1}^{\texttt{\scriptsize N\_test}} (\mathbf{R}^x_{ij})^2, \\
L^v &= \frac{1}{\texttt{\scriptsize N\_test}} \sum_{i=1}^{\texttt{\scriptsize N\_elem}} \sum_{j=1}^{\texttt{\scriptsize N\_test}} (\mathbf{R}^y_{ij})^2, \\
L^c &= \frac{1}{\texttt{\scriptsize N\_test}} \sum_{i=1}^{\texttt{\scriptsize N\_elem}} \sum_{j=1}^{\texttt{\scriptsize N\_test}} (\mathbf{R}^{\text{div}}_{ij})^2, \\
L^{NSE}_{v}(W,b) &= \alpha L^u + \beta L^v + \gamma L^c, \\
L^{NSE}_{\text{VPINN}} &= L^{NSE}_{v} + \tau_D L_d^{NSE} + \tau_N L_n^{NSE}.
\end{aligned}
\end{equation*}
These final equations compute the residual losses for each component (x-momentum, y-momentum, and continuity) and combine them into the variational loss $L^{NSE}_{v}$. The total loss $L^{NSE}_{\text{VPINN}}$ includes the variational loss and the boundary losses $L_d^{NSE}$ (Dirichlet) and $L_n^{NSE}$ (Neumann), weighted by factors $\tau_D$ and $\tau_N$. The weights $\alpha$, $\beta$, and $\gamma$ allow for fine-tuning the relative importance of each component in the overall loss function. The tensor-based loss computation for the Navier-Stokes equations is illustrated in Algorithm~1~(refer Appendix). For further information on the assembly of tensors and the forcing matrices, refer to FastVPINNs paper~\cite{anandh2024fastvpinns}.

\section{Numerical Results}\label{sec:numerical_results}
In the following section, we present a comprehensive validation of the FastVPINNs framework for applications in fluid flow problems. We begin by testing our code on the inviscid Burgers' equation, and extend the FastVPINNs framework to vector-valued problems. Next, we solve both forward and inverse problems using the incompressible Navier-Stokes equations. We validate the accuracy of the FastVPINNs framework by first solving the Kovasznay flow and comparing the predicted solution with the available exact solution. For the Kovasznay flow, we also compare the accuracy and efficiency of FastVPINNs againts NSFnets~\cite{jin2021nsfnets}, which we consider as a benchmark. We also perform a grid-convergence study for Kovasznay flow, showing the effect of element size on the accuracy of the results. Next, we apply the FastVPINNs framework to three canonical examples in fluid flow problems:  lid-driven cavity flow, flow through a rectangular channel and flow past a backward-facing step. 
We also demonstrate the applicability of FastVPINNs to laminar boundary flows, by solving the Falkner-Skan boundary layer problem and comparing our results with those established in literature~\cite{eivazi2022physics}. As described earlier, the FastVPINNs framework can handle complex meshes with skewed quadrilateral elements. To highlight this, we solve the flow past a cylinder problem, demonstrating the flexibility and robustness of FastVPINNs in dealing with non-trivial domain discretizations. Finally, we end our discussion by illustrating the application of our method to inverse problems by predicting the Reynolds number while solving for the flow past a backward-facing step.

The FastVPINNs library~\cite{anandh2024fastvpinnsjoss} has been written with TensorFlow version 2.0~\cite{tensorflow2015-whitepaper}. For the following examples, the test functions are of the form 
\begin{equation*}
    v_k = P_{k+1} - P_{k-1},
\end{equation*}
where, $v_k$ is a polynomial of the $k^{th}$ order and $P_k$ is the $k^{th}$ order Legendre polynomial. For numerical quadrature, the Gauss-Lobatto-Legendre method is utilized. A fully-connected neural network with \texttt{tanh} activation function was used for all experiments, optimized using the Adam optimizer~\cite{kingma2014adam}. The relative $l^2$ error is defined as follows:
\begin{equation*}
L^2_{rel}(u) = \frac{\|u - u_{\text{ref}}\|_2}{\|u_{\text{ref}}\|_2}, 
\end{equation*}
where $\|\cdot\|_2$ is the $l^2$ norm, $u$ is the predicted solution and $u_{\text{ref}}$ is the reference solution. For the benefit of reproducibility of our results, we mention some additional specifications used to run the experiments in Table~\ref{tab:specifications_1}.
\begin{table}[ht]
\begin{tabular}{|c|c|}
\hline
GPU      & NVIDIA A6000                        \\  \hline
CPU      & \begin{tabular}[c]{@{}c@{}}AMD Ryzen Threadripper 3960X \\ 24-Core\end{tabular} \\ \hline
CUDA     & 11.8                                                                            \\ \hline
cuDNN    & 8.6                                                                             \\ \hline
Datatype & \texttt{tf.float32}                                                                      \\ \hline
\end{tabular}

\caption{Specifications used for numerical experiments}
\label{tab:specifications_1}
\end{table}

\subsection{Burgers' Equation}\label{sec:burgers}
\begin{figure*}[t!]
    \centering
    \includegraphics[width=0.7\textwidth]{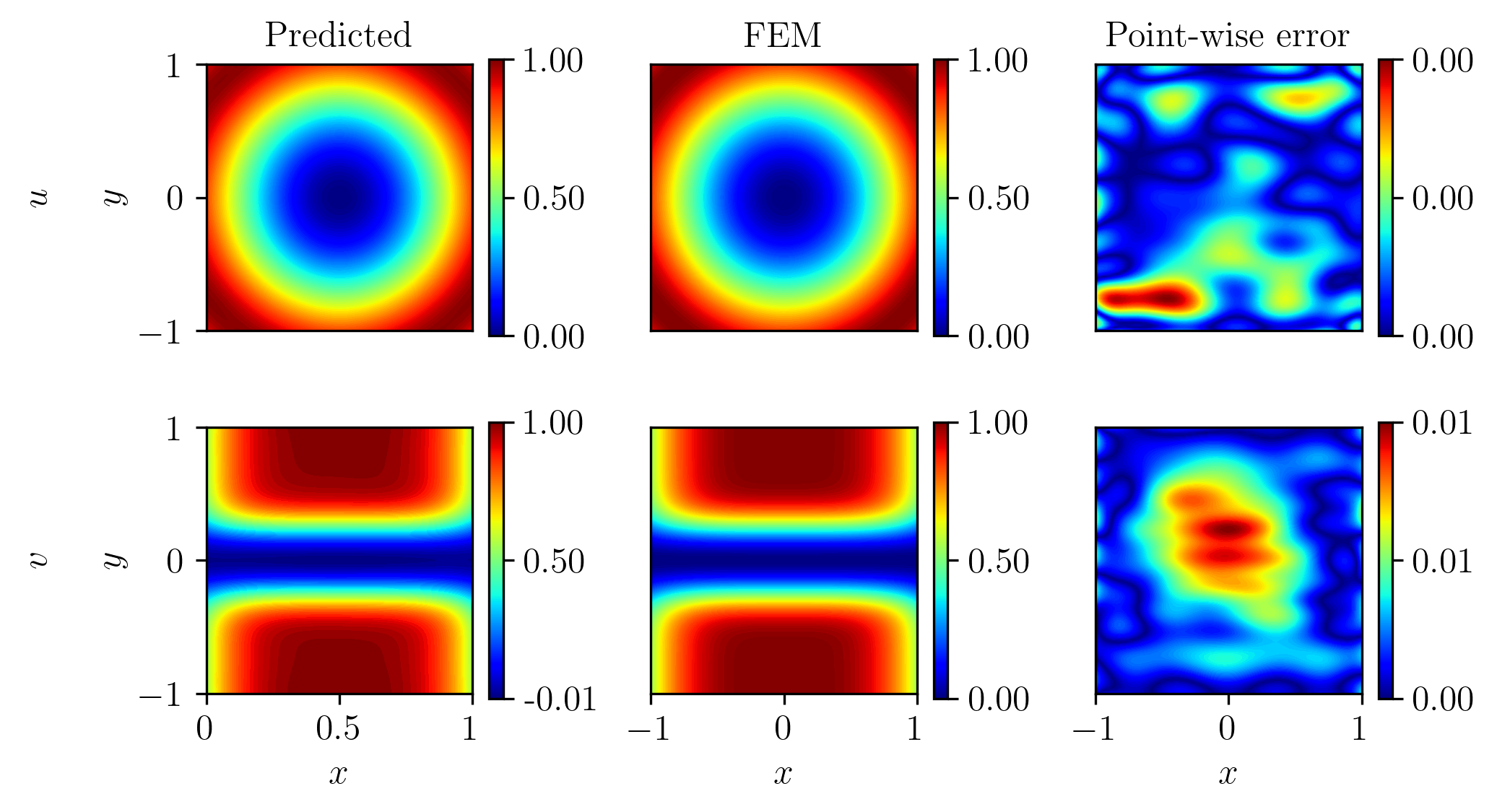}
    \caption{{\bf Solution of Burgers' equation}: Solution predicted by FastVPINNs, exact solution and point-wise errors for $u$ and $v$ respectively.}
    \label{fig:burgers_solution}
\end{figure*}

We begin our discussion on numerical results by solving the 2D, stationary, viscous Burgers' equation, shown in Eq.~\eqref{eqn:burgers2d}. We employ the method of manufactured solutions to generate an exact solution, defined as:
\begin{align}
    \begin{split}
    u &= \sin(x^2 + y^2), \\
    v &= \cos(x^2) \tanh(8y^2).
    \end{split}
    \label{eq:burgers_exact}
\end{align}
The right-hand side terms ($f_x, f_y$) are derived by substituting the values in \eqref{eq:burgers_exact} in Eq.~\eqref{eqn:burgers2d}, with the viscosity coefficient $\nu$ set to 1. The computational domain, $\Omega$ is given by $[-1, 1]^2$, discretized into 8 elements each in the x- and y-dimensions. For this problem, we use a neural network with 3 hidden layers containing 30 neurons each. The network is trained for 20,000 epochs using a constant learning rate of 0.001. To compute the loss, we utilize 25 test functions and 100 quadrature points per element, resulting in a total of 6400 quadrature points. The boundary loss is computed on 800 points, sampled across the boundary of the domain. Fig.~\ref{fig:burgers_solution} shows the exact solution, the solution predicted by FastVPINNs, and the point-wise error for both $u$ and $v$ components. FastVPINNs is able to achieve $L^2_{rel}$ errors of $1.9 \times 10^{-3}$ and $6.1 \times 10^{-3}$ for $u$ and $v$, respectively. Moreover, the mean training time was $2.4ms$ per epoch, highlighting the computational efficiency of our method. 

\subsection{Kovasznay Flow}\label{sec:kovasznay}
We now move on to solving the incompressible Navier-Stokes equation, starting with the Kovasznay flow~\cite{Kovasznay_1948}, a well-known analytical solution of the 2D incompressible Navier-Stokes equation. This flow represents a laminar flow regime behind a two-dimensional grid and is characterized by a parameter $\lambda$ that depends on the Reynolds number. The existence of an analytical solution will provide a clear baseline for assessing the accuracy of the proposed method. It will also allow us to compare the accuracy and speed of our method with the state of the art, NSFnets~\cite{jin2021nsfnets}. 
The analytical solution for the Kovasznay flow is given by:
\begin{align*}
    \begin{split}
    u(x,y) &= 1 - e^{\zeta x} \cos(2 \pi y), \\
    v(x,y) &= \frac{\zeta}{2 \pi} e^{\zeta x} \sin(2 \pi y), \\
    p(x,y) &= \frac{1}{2} \left(1 - e^{2 \zeta x}\right),
    \end{split}
    \label{eq:kovasznay}
\end{align*}
where,
\begin{align*}
\zeta &= \frac{1}{2\mu} - \sqrt{\frac{1}{4 \mu^{2}} + 4\pi^{2}}, \quad \mu = \frac{1}{\text{Re}}.
\end{align*}
We solve the Kovasznay flow at $\text{Re} = 40$, on the rectangular domain $[-0.5,1]\times[-0.5,1.5]$. The domain was discretized with 6 cells in the $x$-dimension and 10 cells in the $y$-dimension, such that $\texttt{N\_elem}=60$. In NSFnets, the best solution was obtained using a neural network with 7 hidden layers, each comprising of 100 neurons, and a total of 2601 collocation points and 400 boundary points. For a fair comparison with NSFnets, we use the same network architecture, along with the parameters in Table \ref{tab:kovasznay_params}. The weighting coefficients $\alpha$, $\beta$, and $\gamma$ in Eq.~\eqref{eq:variational_loss_NSE} are all set to 10.
\begin{table}[ht!]
    \centering
    \begin{tabular}{|l|c|c|}
        \hline
        \textbf{Parameter} & \textbf{FastVPINNs} & \textbf{NSFnets}\\
        \hline
        Number of elements & 60 (6 $\times$ 10) & - \\
        \hline
        Quadrature/Collocation points & 2160 & 2601 \\
        \hline
        Boundary points & 400 & 400 \\
        \hline
        Test functions per element & 16 & - \\
        \hline
        Neural network architecture & 7 hidden layers, & 7 hidden layers, \\
        & 100 neurons each & 100 neurons each \\
        \hline
        Training epochs & 40,000 & 40,000 \\
        \hline
    \end{tabular}
    \caption{Kovasznay flow: Parameters used for FastVPINNs and NSFnets.}
    \label{tab:kovasznay_params}
\end{table}

\begin{figure*}[ht!]
     \centering
\includegraphics[width=0.7\textwidth]{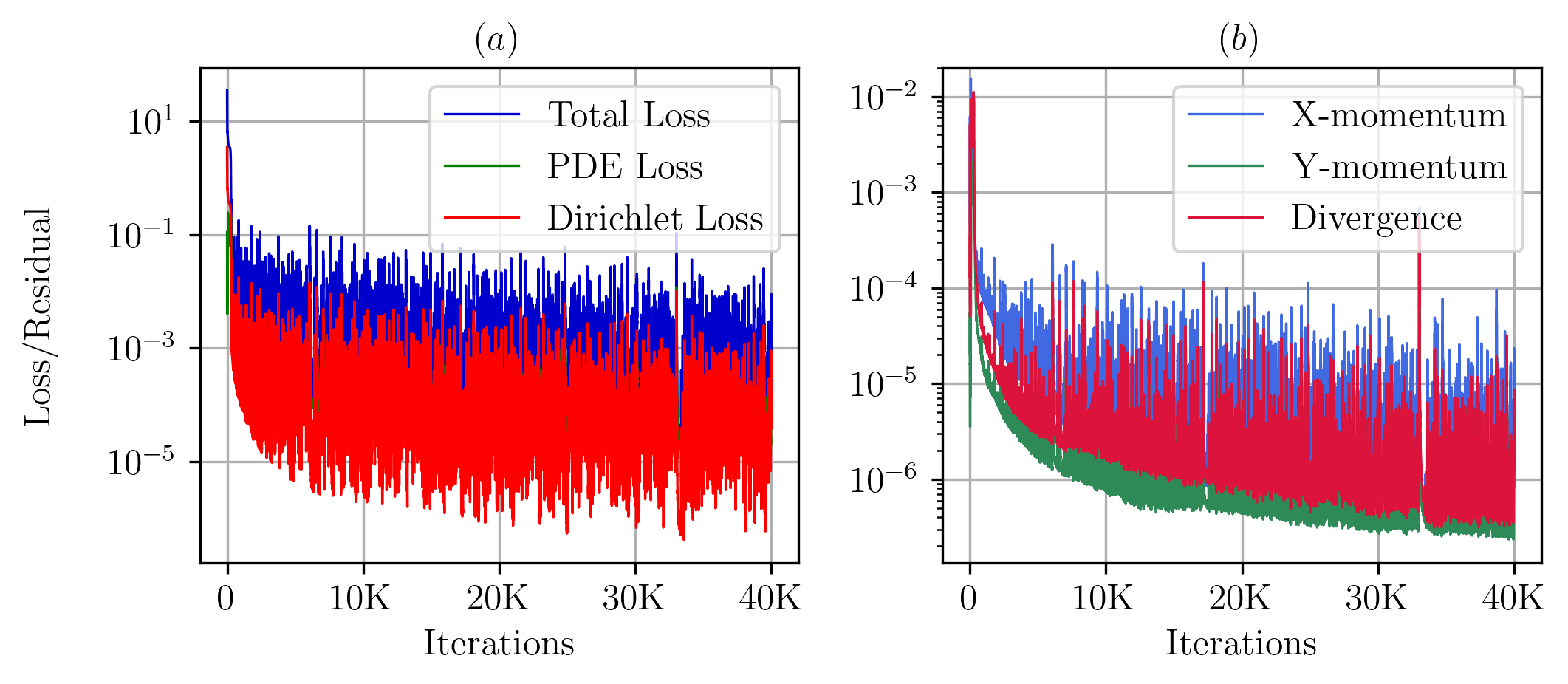}
     \caption{{\bf Training for Kovasznay flow: }(a) PDE loss, Dirichlet boundary loss and total loss during training (b) PDE loss split into its components: x-momentum residual, y-momentum residual and divergence loss.}
     \label{fig:kovasznay_loss}
 \end{figure*}

\begin{figure*}[ht!]
     \centering
     \includegraphics[width=0.7\textwidth]{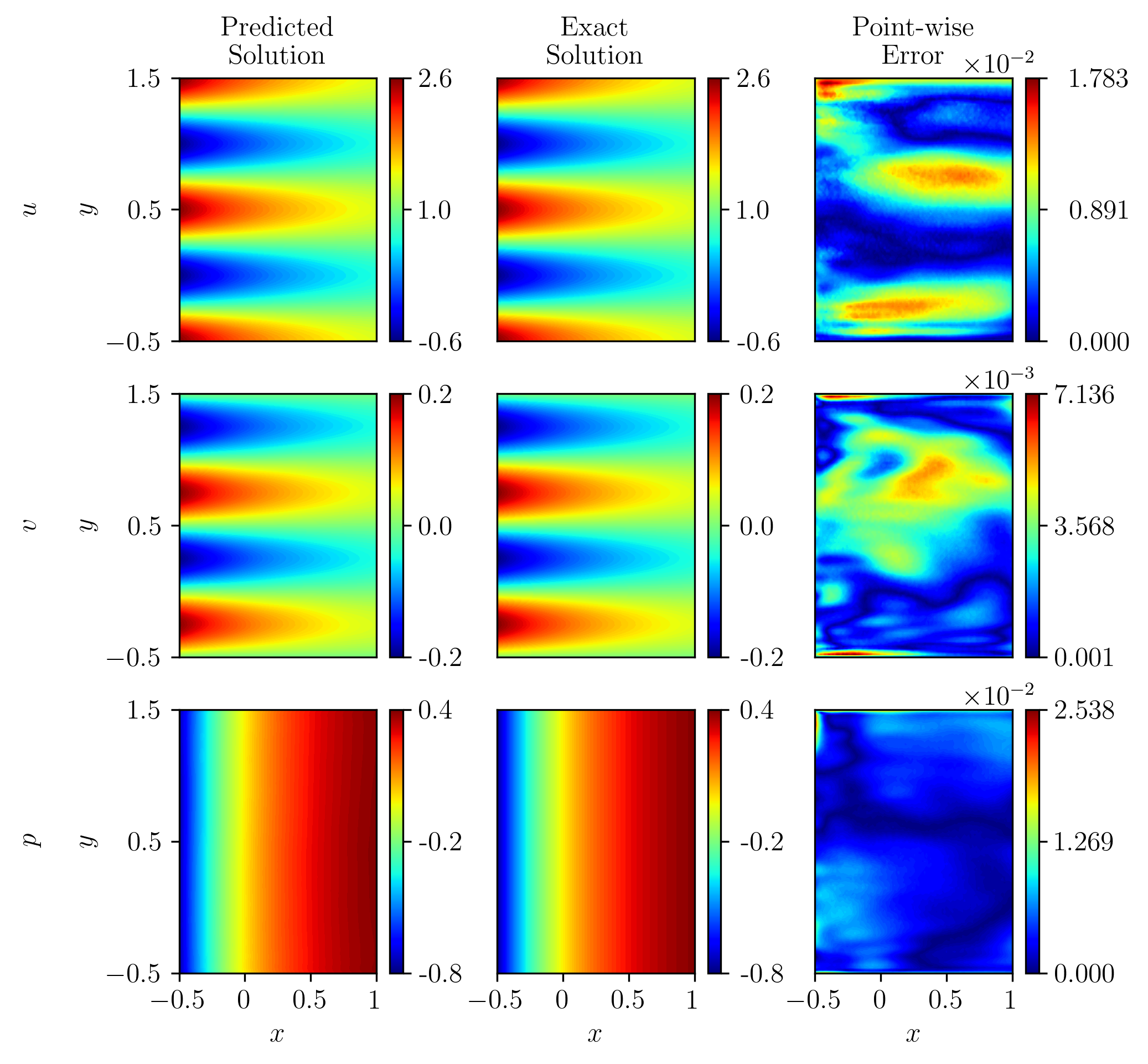}
     \caption{{\bf Solution of Kovasznay flow for $\text{Re}=40$}: Solution predicted by FastVPINNs, exact solution and point-wise errors for $u$, $v$ and $p$ respectively.}
     \label{fig:kovasznay_solution}
 \end{figure*}

As with NSFnets, we train FastVPINNs for 40,000 iterations, resulting in the loss functions shown in Fig.~\ref{fig:kovasznay_loss}. The predicted solution and point-wise errors for each component are shown in Fig.~\ref{fig:kovasznay_solution}. A comparison between NSFnets and FastVPINNs, in terms of accruacy and speed, is shown in Table \ref{tab:kovasznay_accuracy}. The table has been compiled using the mean result and standard deviation of 5 runs with independent initial network parameters for both NSFnets and FastVPINNs. We observe that FastVPINNs matches NSFnets in terms of accuracy, while only requiring about 40\% of the training time. A similar comparison for a network with 4 hidden layers with 50 neurons each can be found in Table~\ref{tab:kovasznay_accuracy_50x4}~(in appendix). 
\renewcommand{\arraystretch}{1.2}
\begin{table}[ht!]
\centering
\begin{tabular}{|c|c|c|}
\hline
\textbf{Metric} & \textbf{FastVPINNs} & \textbf{NSFnets} \\ \hline
\textbf{$L^2_{rel} (u)$} & $(2.64 \pm 1.22) \times 10^{-3}$ & $(2.60 \pm 0.50) \times 10^{-3}$ \\ \hline
\textbf{$L^2_{rel} (v)$} & $(3.10 \pm 2.67) \times 10^{-2}$ & $(2.60 \pm 0.66) \times 10^{-2}$ \\ \hline
\textbf{$L^2_{rel} (p)$} & $(7.03 \pm 2.39) \times 10^{-3}$ & $(9.40 \pm 2.30) \times 10^{-3}$ \\ \hline
\begin{tabular}[c]{@{}c@{}}Training Time for\\ 1K Iterations (s)\end{tabular} & \textbf{7.51} & 20.351 \\ \hline
\end{tabular}
\caption{Comparison of accuracy and training time between FastVPINNs and NSFnets for Kovasznay flow at $\text{Re}=40$.}
\label{tab:kovasznay_accuracy}
\end{table}

\subsubsection*{Grid Convergence for Kovasznay Flow}
\begin{figure}[t!]
     \centering
\includegraphics[width=0.38\textwidth]{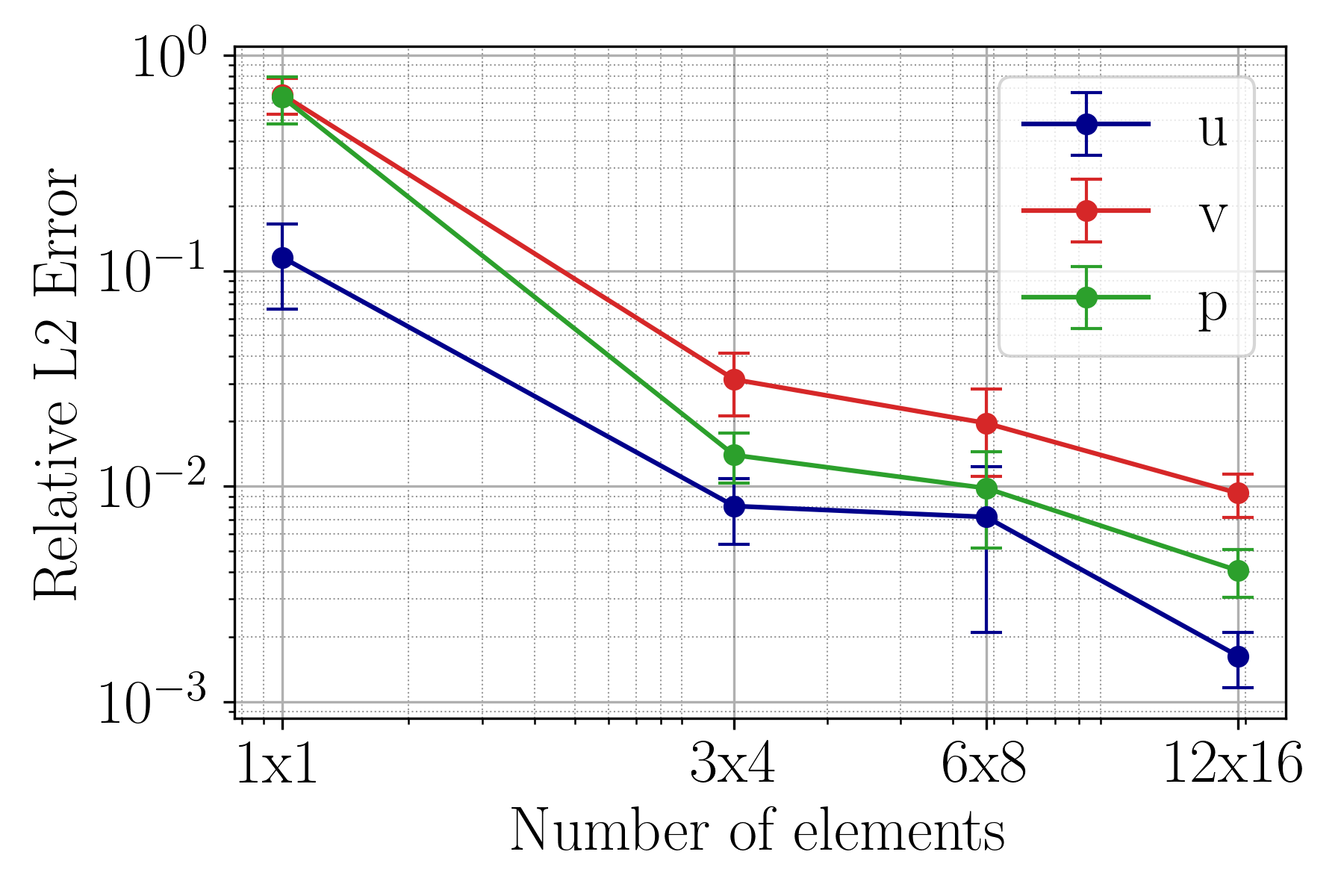}
     \caption{{\bf Kovasznay Flow: } Relative errors of $u$, $v$ and $p$ for various grid sizes}
     \label{fig:kovasznay_convergence}
 \end{figure}

To demonstrate the convergence properties of hp-VPINNs with an increasing number of elements, we perform a grid convergence study for the Kovasznay flow problem. The number of quadrature points per element is fixed at 36. The number of test functions, boundary points, and neural network architecture remain the same as shown in Table~\ref{tab:kovasznay_params}. To ensure proper training of the network as the number of elements increases, we adjust the number of training iterations according to the following relation:
\begin{equation*}
    \text{Number of training iterations} = \kappa + \xi \times \texttt{N\_quad}*\texttt{N\_elem}
\end{equation*}
where $\kappa = 6000$ represents the base number of epochs, and $\xi = 6$ is a scaling factor that accounts for the increase in quadrature points within the domain. Table~\ref{tab:grid_convergence} and Fig.~\ref{fig:kovasznay_convergence} show the relative $L^2$ error of $u$, $v$, and $p$. We observe that with sufficient training, the test error decreases as the number of elements in the domain increases.


\begin{table}[ht]
\centering
\begin{tabular}{|c|c|c|c|c|}
\hline
\textbf{Grid Size} & \textbf{Epochs} & \textbf{$\boldsymbol{L^2_\text{rel}}~\boldsymbol{(u)}$} & \textbf{$\boldsymbol{L^2_\text{rel}}~\boldsymbol{(v)}$} & \textbf{$\boldsymbol{L^2_\text{rel}}~\boldsymbol{(p)}$} \\
\hline
1$\times$1  & 6,600 & $(1.15 \pm 0.49)$ & $(6.55 \pm 1.24)$ & $(6.37 \pm 1.57)$ \\
 &  & $\times 10^{-1}$ & $\times 10^{-1}$ & $\times 10^{-1}$ \\
\hline
3$\times$4  & 13,200 & $(8.10 \pm 2.72)$ & $(3.13 \pm 1.01)$ & $(1.40 \pm 0.37)$ \\
 &  & $\times 10^{-3}$ & $\times 10^{-2}$ & $\times 10^{-2}$ \\
\hline
6$\times$8  & 34,800 & $(7.21 \pm 5.11)$ & $(1.96 \pm 0.85)$ & $(9.80 \pm 4.64)$ \\
 &  & $\times 10^{-3}$ & $\times 10^{-2}$ & $\times 10^{-3}$ \\
\hline
12$\times$16 & 121,200 & $(1.63 \pm 0.47)$ & $(9.30 \pm 2.10)$ & $(4.07 \pm 1.01)$ \\
 &  & $\times 10^{-3}$ & $\times 10^{-3}$ & $\times 10^{-3}$ \\
\hline
\end{tabular}
\caption{Grid convergence study for Kovasznay flow}
\label{tab:grid_convergence}
\end{table}

\subsection{Lid Driven Cavity Flow}\label{sec:ldc}
\begin{figure*}
\includegraphics[width=0.7\textwidth]{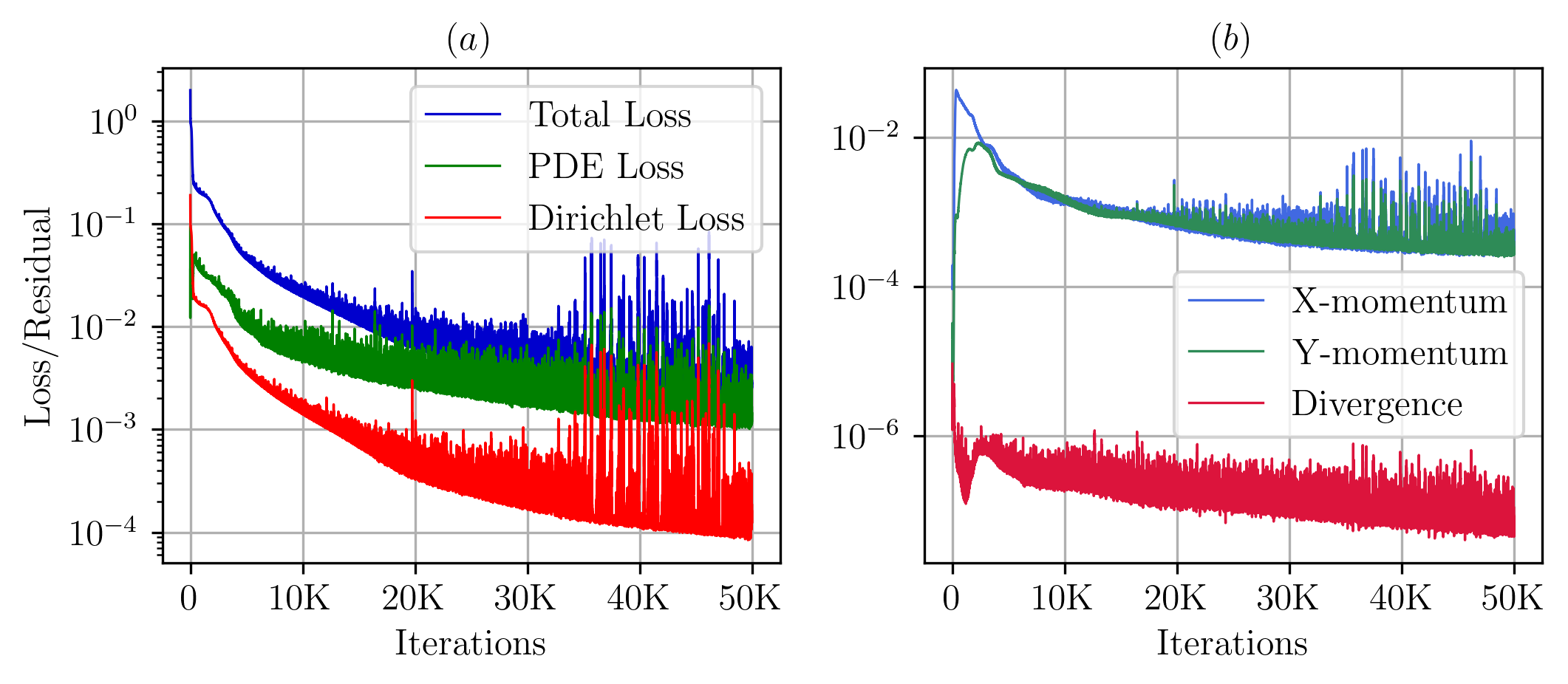}
     \caption{{\bf Training for lid-driven cavity flow at Re = 1}: a) PDE loss, Dirichlet boundary loss and total loss during training (b) PDE loss split into its components: x-momentum residual, y-momentum residual and divergence loss.}
     \label{fig:ldc_loss}
 \end{figure*}

\begin{figure*}[t!]
    \centering
    \includegraphics[width=0.7\textwidth]{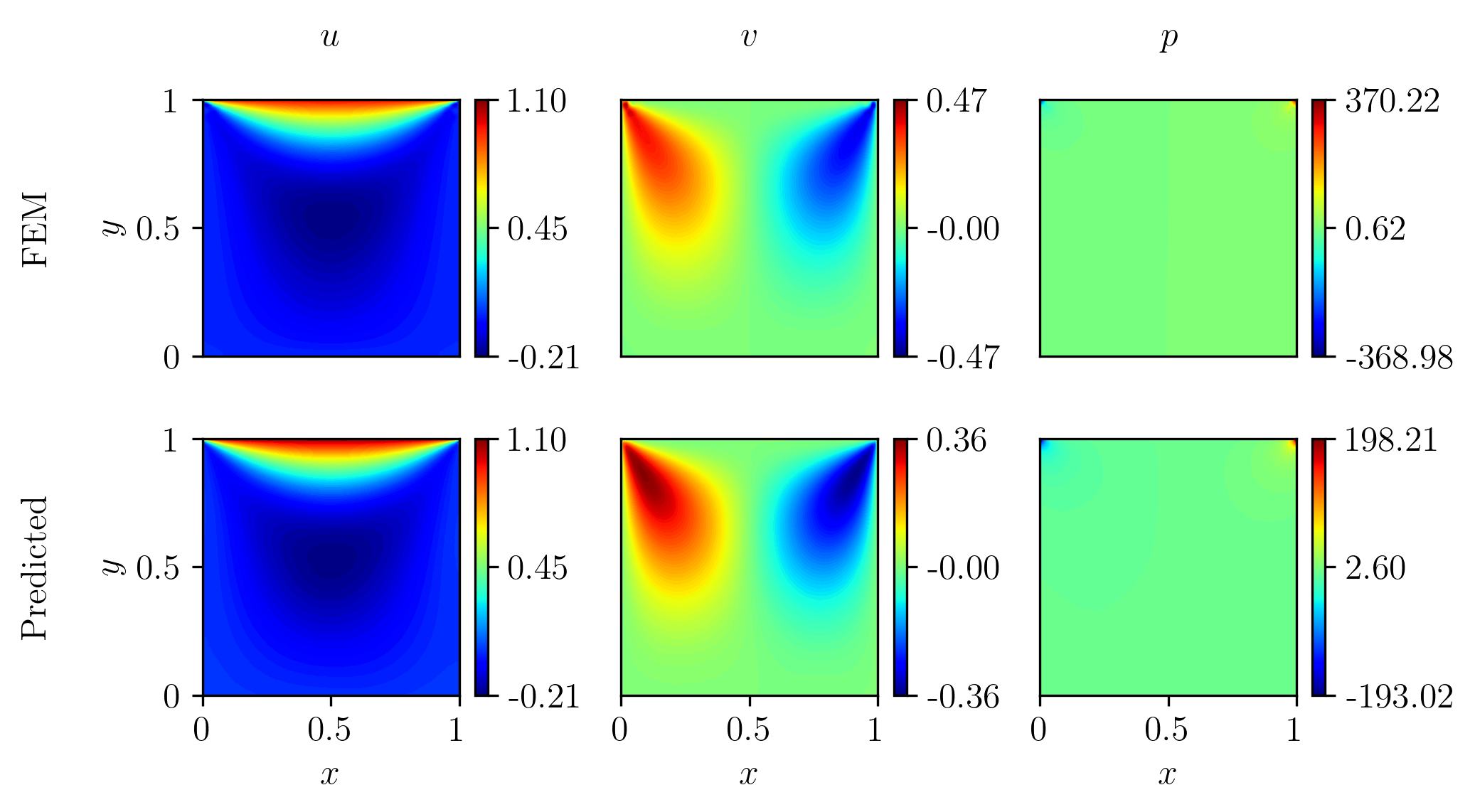}
    \caption{{\bf Solution of lid-driven cavity flow at Re = 1}: Solution predicted by FastVPINNs and FEM solution for $u$, $v$ and $p$ respectively.}
    \label{fig:lidddriven_solution}
\end{figure*}

\begin{figure}[t!]
    \centering
\includegraphics[width=0.5\textwidth]{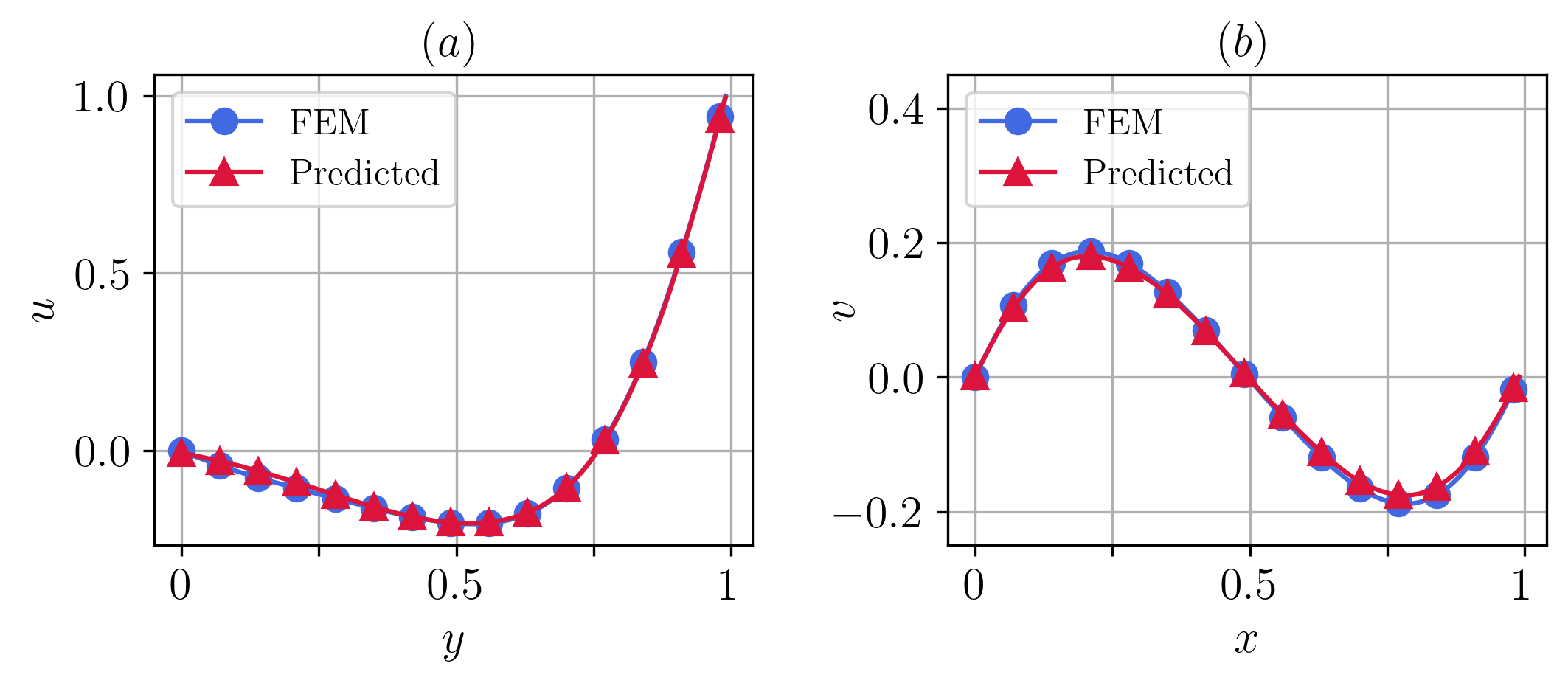}
    \caption{{\bf Mid-line solution of lid-driven cavity flow at Re = 1}: (a) Comparison between FEM and FastVPINNs solution for $u$ at $x=0.5$. (b) Comparison between FEM and FastVPINNs solution for $v$ at $y=0.5$.}
    \label{fig:liddriven_midline}
\end{figure}

Following our investigation of the Kovasznay flow, we now turn our attention to another classic benchmark problem in fluid dynamics: the lid-driven cavity flow. We discretize the unit square domain using 8 quadrilateral elements in each dimension, with the parameters shown in Table~\ref{tab:ldc_params}. 
\begin{table}[H]
    \centering
    \begin{tabular}{|c|c|}
        \hline
        Total number of elements, $\texttt{N\_elem}$ & 64\\
        \hline
        Quadrature points per element, $\texttt{N\_quad}$& 100\\
        \hline
        Total number of quadrature points & 6400\\
        \hline
        Number of test functions per element, $\texttt{N\_test}$ & 36\\
        \hline
        Number of boundary points, $N_D$ & 800 \\
        \hline
    \end{tabular}
    \caption{FastVPINNs parameters for lid-driven cavity flow.}
    \label{tab:ldc_params}
\end{table}
For lid-driven cavity flow, a neural network architecture consisting of six hidden layers, each containing 20 neurons has been used. The network was trained for 50,000 epochs, as shown in Fig.~\ref{fig:ldc_loss}, with an initial learning rate of 0.0013. This learning rate was decayed using an exponential learning rate scheduler, with a decay rate of 0.99 every 1000 steps. we set the weighting coefficients in Eq.~\eqref{eq:variational_loss_NSE} as $\alpha = 1$, $\beta = 1$, and $\gamma = 10^4$. This choice emphasizes the importance of the continuity equation in the loss function. Fig.~\ref{fig:lidddriven_solution} presents the comparison between the FEM solution computed using ParMooN~\cite{wilbrandt2017parmoon} and the solution predicted by the neural network. Fig.~\ref{fig:liddriven_midline} illustrates the midline velocity comparison between the FEM values and the predicted values from the neural network and we observe that our neural network solution is in good agreement with the FEM solution. 

\subsection{Flow through channel}\label{sec:channel}
\begin{figure*}[ht!]
    \centering
    \includegraphics[width=0.75\textwidth]{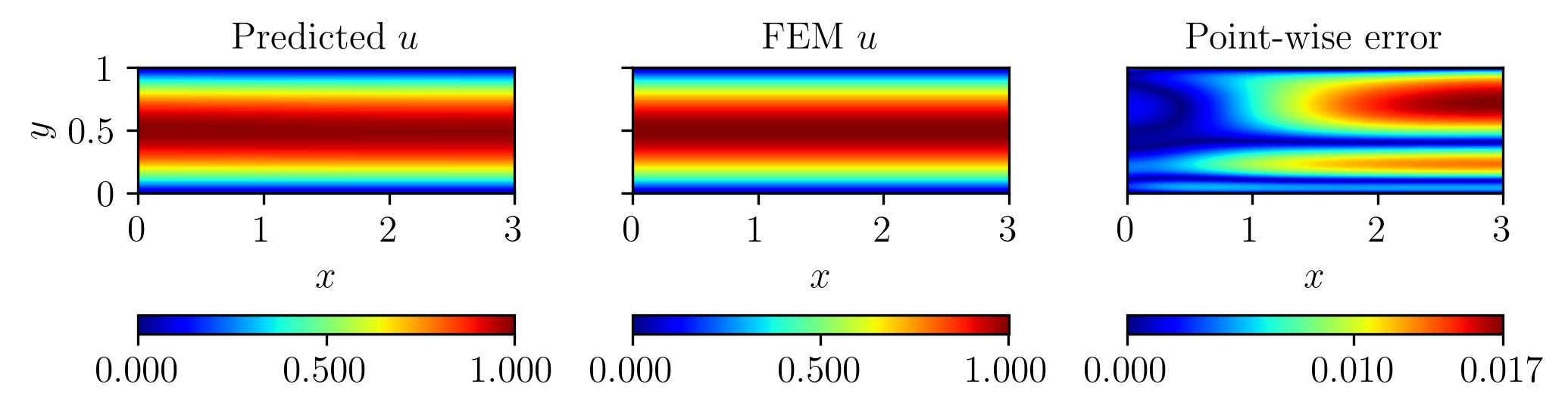}
    \caption{{\bf Solution of Flow through Channel at Re = 1}: Solution predicted by FastVPINNs, FEM solution and point-wise errors for $u$, $v$ and $p$.}
    \label{fig:flow_channel}
\end{figure*}

Next, we solve for the flow through a rectangular channel, to validate our implementation of Neumann boundary conditions. This case study involves fluid flow in the rectangular domain $[0,3]\times[0,1]$, discretized using a hundred regular quadrilateral elements. The inlet velocity profile is defined as 
\begin{equation*}
    u(0,y) = 4y(1-y),
\end{equation*}
achieving a maximum value of 1 at $y=0.5$. We impose the no-slip boundary condition on the top and bottom walls, while the outlet has a zero Neumann boundary condition. A total of 800 boundary points were distributed across the domain boundary to enforce the boundary conditions. The weighting coefficients in Eq.~\eqref{eq:variational_loss_NSE} are set as $\alpha = 10^{-4}$, $\beta = 10^{-4}$, and $\gamma = 10^{4}$. Table~\ref{tab:channel_flow_params} summarizes the key hyperparameters used in this simulation.  A learning rate of $0.001$ was used for the Adam optimizer, and the training was carried out for 30,000 epochs. For a Reynolds number of 1, we obtained a $L^2_{rel}$ error of $1.09\times 10^{-2}$ for the velocity in the $x$-dimension. Fig.~\ref{fig:flow_channel} shows the predicted $u$ velocity and the corresponding point-wise error.

\begin{table}[H]
\centering
\begin{tabular}{|c|c|}
\hline
Neural network architecture                         & \makecell{5 hidden layers, \\ 30 neurons each}              \\ \hline
Number of elements                   & 100 $(20 \times 5)$ \\ \hline
Number of quadrature points per cell & 36                                            \\ \hline
Number of test functions per cell    & 16                                            \\ \hline
\end{tabular} 
\caption{FastVPINNs parameters for flow through channel}
\label{tab:channel_flow_params}
\end{table}

\subsection{Flow past a backward facing step}\label{sec:bfs}
\begin{figure}[ht!]
    \centering
    \includegraphics[width=0.48\textwidth]{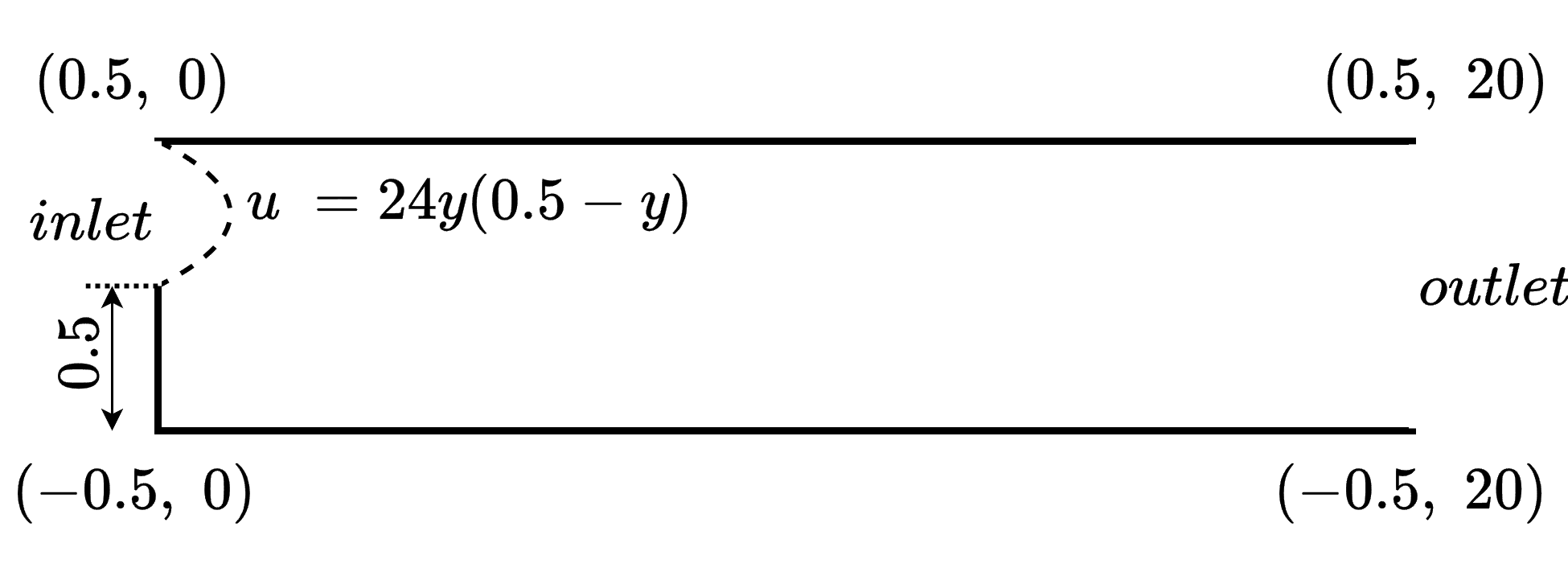}
    \caption{Domain and inlet velocity profile for flow past a backward-facing step}
    \label{fig:fpbfs_mesh}
\end{figure}

\begin{figure*}[ht!]
    \centering
    \includegraphics[width=0.7\textwidth]{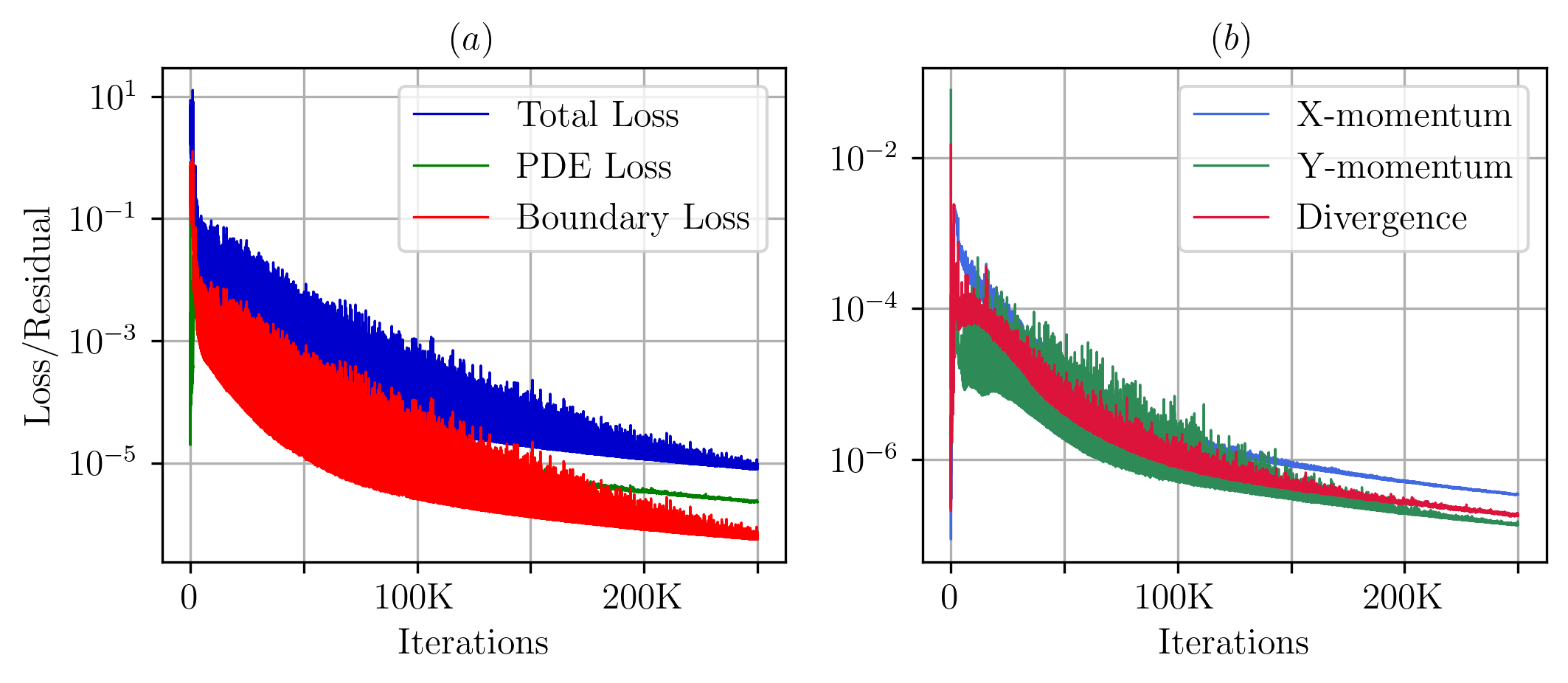}
    \caption{{\bf Solution of flow past a backward-facing step at Re = 200}: a) PDE loss, Dirichlet boundary loss and total loss during training (b) PDE loss split into its components: x-momentum residual, y-momentum residual and divergence loss.}
    \label{fig:backwardstep_loss}
\end{figure*}

\begin{figure*}[t!]
    \centering
\includegraphics[width=0.75\textwidth]{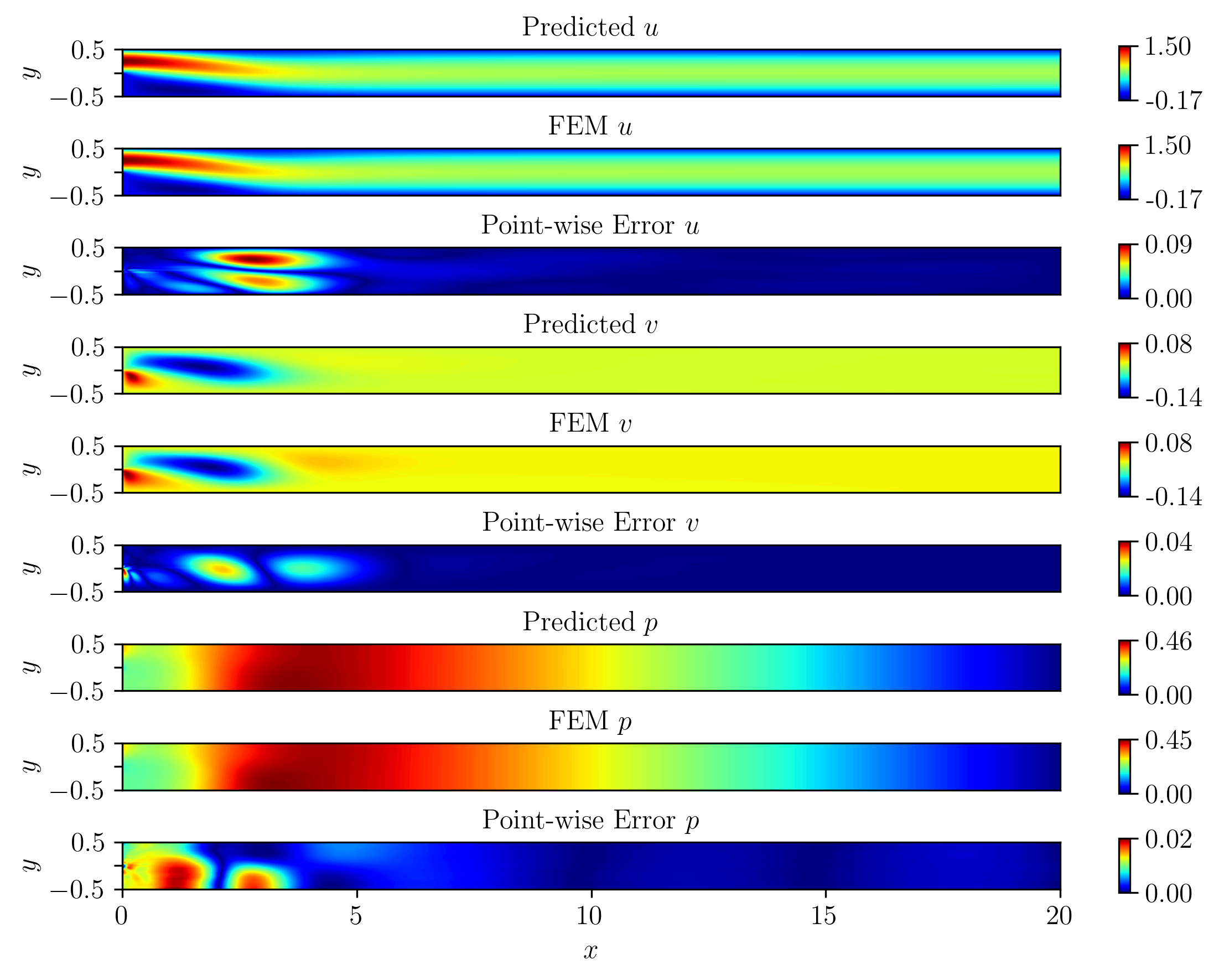}
    \caption{{\bf Solution of flow past a backward-facing step at Re = 200}: Solution predicted by FastVPINNs, FEM solution and point-wise errors for $u$, $v$ and $p$.}
    \label{fig:backwardstep_solution}
\end{figure*}

The backward-facing step problem, a challenging case in fluid dynamics, involves flow through a channel with a sudden expansion. This configuration leads to complex phenomena such as flow separation, reattachment, and recirculation zones. We investigate this problem at a Reynolds number of 200, which presents a significant test for our FastVPINNs framework due to the large domain and varying velocity fields along the channel length. The computational domain, similar to that in Gartling~\cite{gartling1990test}, is illustrated in Fig.~\ref{fig:fpbfs_mesh}. This figure shows both the geometry of the backward-facing step and the inlet velocity profile used in our simulations.
We discretize the domain into 100 elements $(20 \times 5)$, with the computational parameters detailed in Table~\ref{tab:bfs_params}.
\begin{table}[H]
    \centering
    \begin{tabular}{|c|c|}
        \hline
        Total number of elements, $\texttt{N\_elem}$ & 100 $(20 \times 5)$\\
        \hline
        Quadrature points per element, $\texttt{N\_quad}$& 64\\
        \hline
        Total number of quadrature points & 6400\\
        \hline
        Number of test functions per element, $\texttt{N\_test}$ & 25\\
        \hline
        Number of boundary points, $N_D$ & 800 \\
        \hline
    \end{tabular}
    \caption{FastVPINNs parameters for backward-facing step flow.}
    \label{tab:bfs_params}
\end{table}
For this problem, we employ a neural network with eight hidden layers, each containing 50 neurons. The network was trained for 250,000 epochs, as shown in Fig.~\ref{fig:backwardstep_loss}, with an initial learning rate of 0.0015. This learning rate was decayed using an exponential learning rate scheduler, with a decay rate of 0.98 every 1000 steps. The weighting coefficients in Eq.~\eqref{eq:variational_loss_NSE} were all set to 10.
To validate our approach, we compare our results with those obtained from FEM. Table~\ref{tab:backwardstep} presents the $L^2_{rel}$ errors of the velocity components $u$, $v$, and pressure $p$. Fig.~\ref{fig:backwardstep_solution} illustrates the comparison between the predicted solution by neural network,  FEM solution, and the point-wise error for both velocity and pressure components. Our analysis shows that the predicted solution from FastVPINNs is in good agreement with the FEM solution, demonstrating the effectiveness of our framework for this complex flow problem.
\begin{table}[htbp!]
\centering
\begin{tabular}{|c|c|}
\hline
\textbf{Component} & \textbf{Relative $L^2$ Error} \\
\hline
$u$ (x-velocity) & $2.73 \times 10^{-2}$ \\
\hline
$v$ (y-velocity) & $1.72 \times 10^{-1}$ \\
\hline
$p$ (pressure) & $1.33 \times 10^{-2}$ \\
\hline
\end{tabular}
\caption{Relative $L^2$ errors for FastVPINNs solution of flow past backward facing step at Reynolds number of 200 after 250k Training iterations.}
\label{tab:backwardstep}
\end{table}
\subsection{Falkner Skan flow}\label{sec:falkner_skan}
\begin{figure*}[htbp!]
    \centering
    \includegraphics[width=0.7\textwidth]{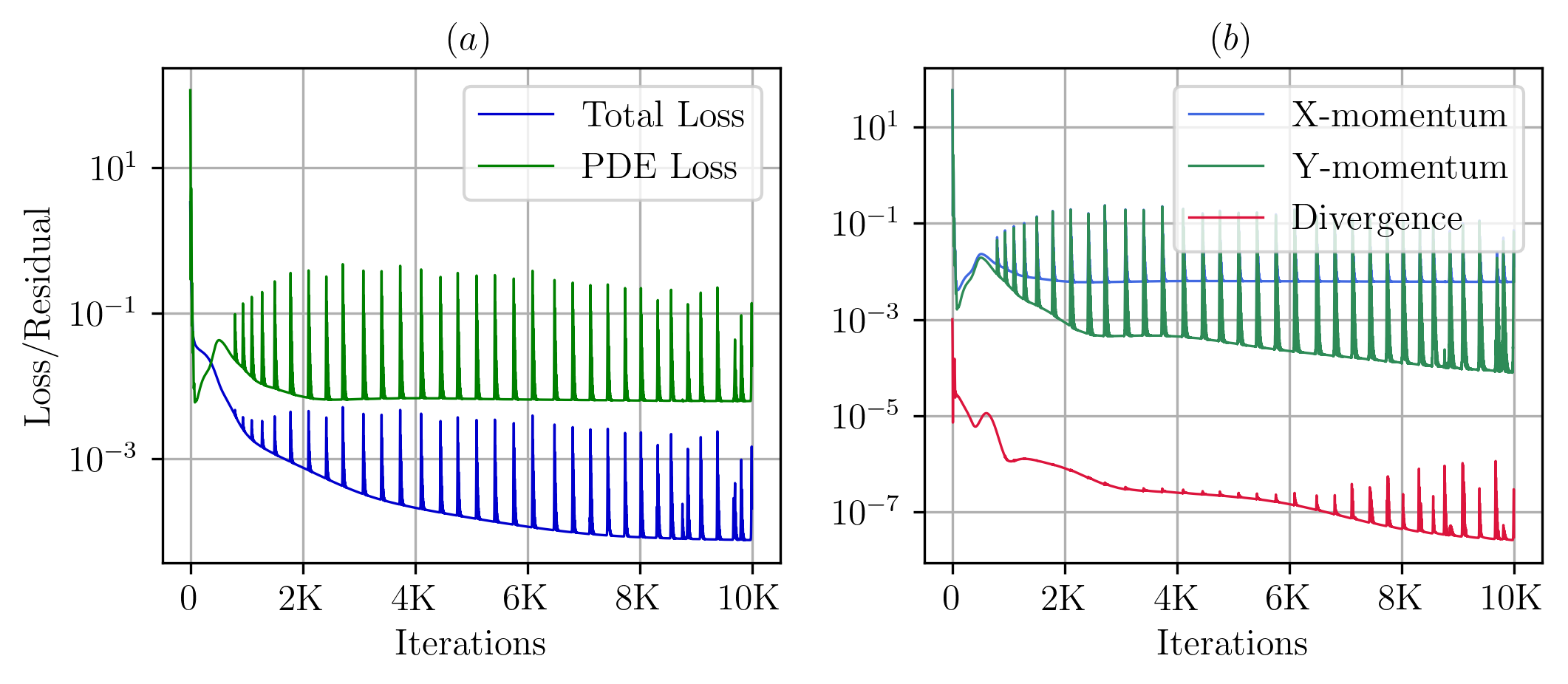}
    \caption{{\bf Training for Falkner-Skan boundary layer}: a) PDE loss, Dirichlet boundary loss and total loss during training (b) PDE loss split
into its components: x-momentum residual, y-momentum residual and divergence loss.}
    \label{fig:falkner_skan_loss}
\end{figure*}

\begin{figure*}[htbp!]
    \centering
    \includegraphics[width=0.7\textwidth]{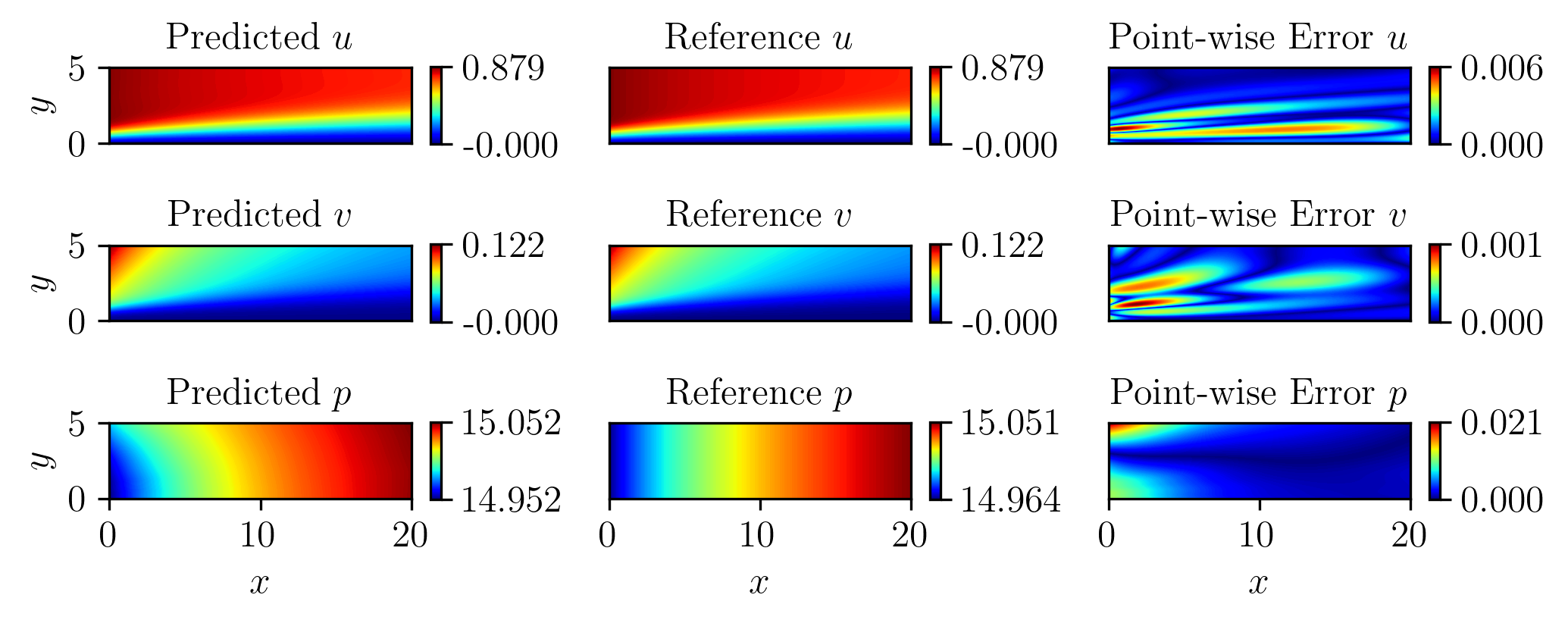}
    \caption{{\bf Solution of Falkner-Skan boundary layer}: Solution predicted by FastVPINNs, reference solution and point-wise errors for $u$, $v$ and $p$.}
    \label{fig:falkner-skan-solution}
\end{figure*}

Following our investigation of classical fluid dynamics problems, we now examine the Falkner-Skan boundary layer problem~\cite{falkneb1931lxxxv}, an important case study in laminar flow analysis. This problem describes the velocity profile of a steady, incompressible boundary layer flow in the laminar region. We have chosen this problem to demonstrate the performance of FastVPINNs compared to traditional PINNs implementations. For our study, we adopted the problem parameters from Eivazi et al.~\cite{eivazi2022physics}, considering a Reynolds number of 100 with $m=-0.08$, resulting in a wedge angle $\beta = -0.1988$. The reference solution and the PINNs code were obtained from the official GitHub repository~\cite{kth_flowai_pinns_2023}. The computational domain spans $[0, 20]$ in the $x$-dimension and $[0, 5]$ in the $y$-dimension.
Table~\ref{tab:falkner_skan_params} summarizes the key parameters used in our FastVPINNs simulation and the comparison PINNs implementation. Both approaches used the Adam optimizer to train for 10,000 epochs. We report the mean results for 10 independent runs. Fig.~\ref{fig:falkner_skan_loss} illustrates the progression of the training, showing the total loss and its components over the training iterations. The weighting coefficients $\alpha$, $\beta$, and $\gamma$ in Eq.~\eqref{eq:variational_loss_NSE} have been set to 1.
\begin{table}[ht!]
    \centering
    \begin{tabular}{|c|c|c|}
        \hline
        \textbf{Parameter} & \textbf{FastVPINNs} & \textbf{PINNs~\cite{eivazi2022physics}} \\
        \hline
        Number of elements & 100 (20 $\times$ 5) & N/A \\
        \hline
        Quadrature/Collocation points & 3600 & 3618 \\
        \hline
        Boundary points & 500 & 500 \\
        \hline
        Test functions per element & 16 & N/A \\
        \hline
        Neural network architecture & 6 hidden layers, & 6 hidden layers, \\
        & 20 neurons each & 20 neurons each \\
        \hline
        Training epochs & 10,000 & 10,000 \\
        \hline
    \end{tabular}
    \caption{Falkner-Skan problem: Parameters used for FastVPINNs and PINNs.}
    \label{tab:falkner_skan_params}
\end{table}
The results presented in Fig.~\ref{fig:falkner-skan-solution}, demonstrate that FastVPINNs achieved velocity solutions comparable to the PINNs code in terms of the relative $L^2$ error.
Table~\ref{tab:fs} details the performance comparison between FastVPINNs and PINNs.
\begin{table}[ht!]
    \centering
    \begin{tabular}{|c|c|c|}
        \hline
        & \textbf{FastVPINNs} & \textbf{PINNs} \\
        \hline
        \textbf{$L^2_{rel}(u)$} & $(3.40 \pm 0.76) \times 10^{-3}$ & $(5.87 \pm 1.49) \times 10^{-3}$ \\
        \hline
        \textbf{$L^2_{rel}(v)$} & $(9.51 \pm 3.12) \times 10^{-3}$ & $(1.69 \pm 0.66) \times 10^{-2}$ \\
        \hline
        \textbf{$L^2_{rel}(p)$} & $(2.68 \pm 0.04) \times 10^{-4}$ & $(7.24 \pm 2.61) \times 10^{-5}$ \\
        \hline
        Time (s) & 7.43 & 16.67 \\
        \hline
    \end{tabular}
    \caption{Performance comparison between FastVPINNs and PINNs for the Falkner-Skan problem. Time (s) represents the duration for 1000 training epochs.}
    \label{tab:fs}
\end{table}
Notably, we were able to achieve a significant speed-up with FastVPINNs. We calculated the average training time for a thousand iterations and found that FastVPINNs achieved a speed-up of 2.24 times when compared with the PINNs implementation. Specifically, the PINNs code from~\cite{kth_flowai_pinns_2023} required 16.67 seconds per 1000 training iterations, while FastVPINNs completed the same number of iterations in just 7.43 seconds.
\subsection{Flow past a cylinder}\label{sec:fpc}
\begin{figure}[t!]
    \centering
    \includegraphics[width=0.33\textwidth]{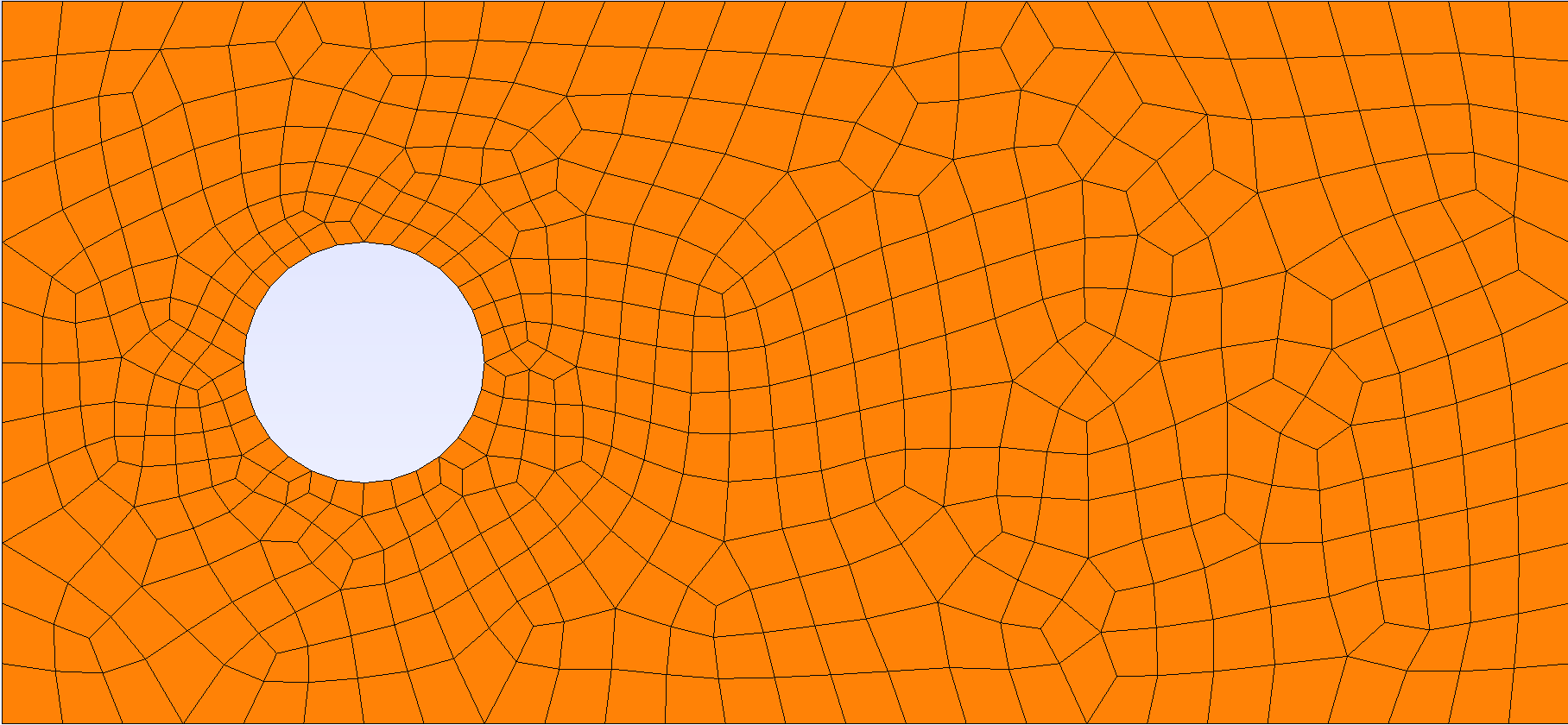}
    \caption{Computational mesh for the flow past cylinder problem.}
    \label{fig:fpc_mesh}
\end{figure}

\begin{figure*}[tbp!]
    \centering
    \includegraphics[width=0.7\textwidth]{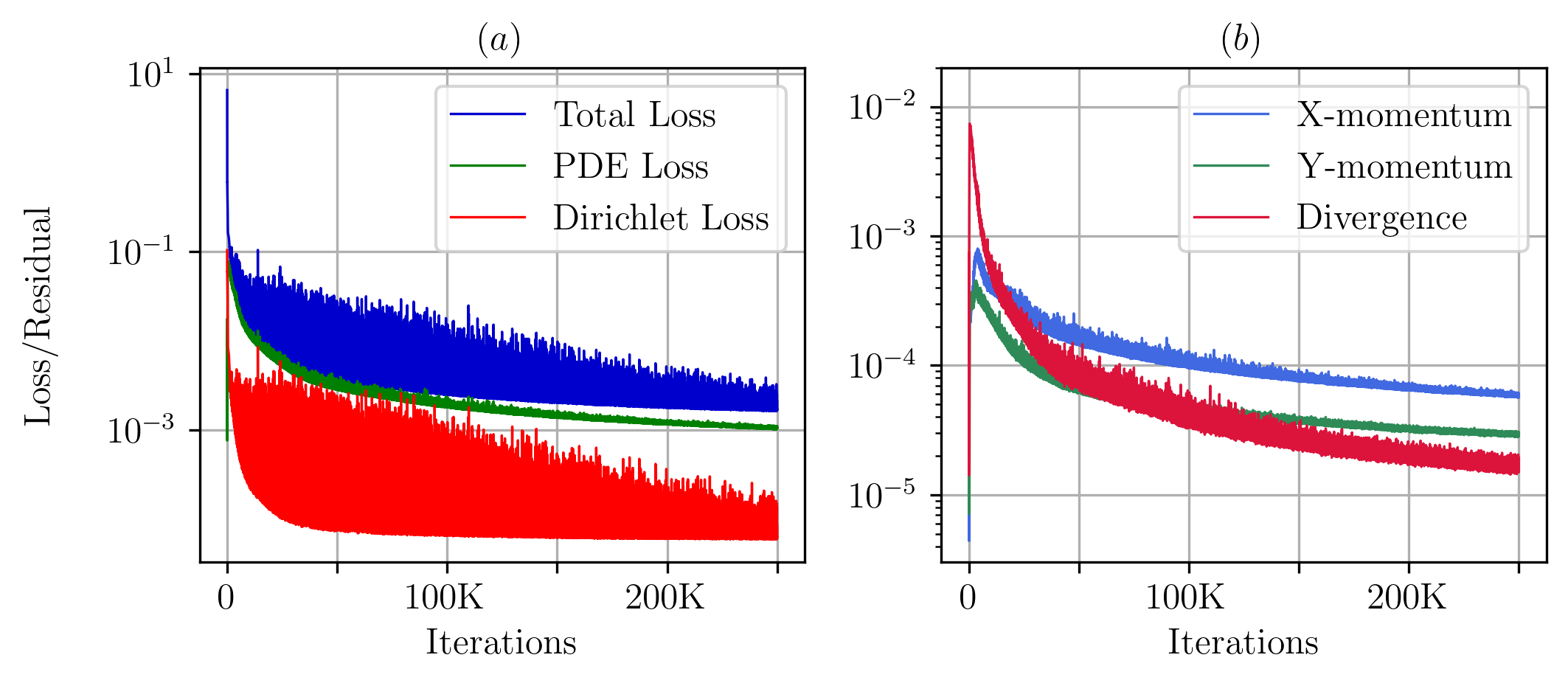}
    \caption{{\bf Training for flow past a cylinder at Re = 50}: (a) PDE loss, Dirichlet boundary loss and total loss during training (b) PDE Loss split into its components - momentum residual in x-direction, momentum residual in y-direction and divergence loss.}
    \label{fig:fpc_loss}
\end{figure*}

\begin{figure*}[t!]
    \centering
    \includegraphics[width=0.79\textwidth]{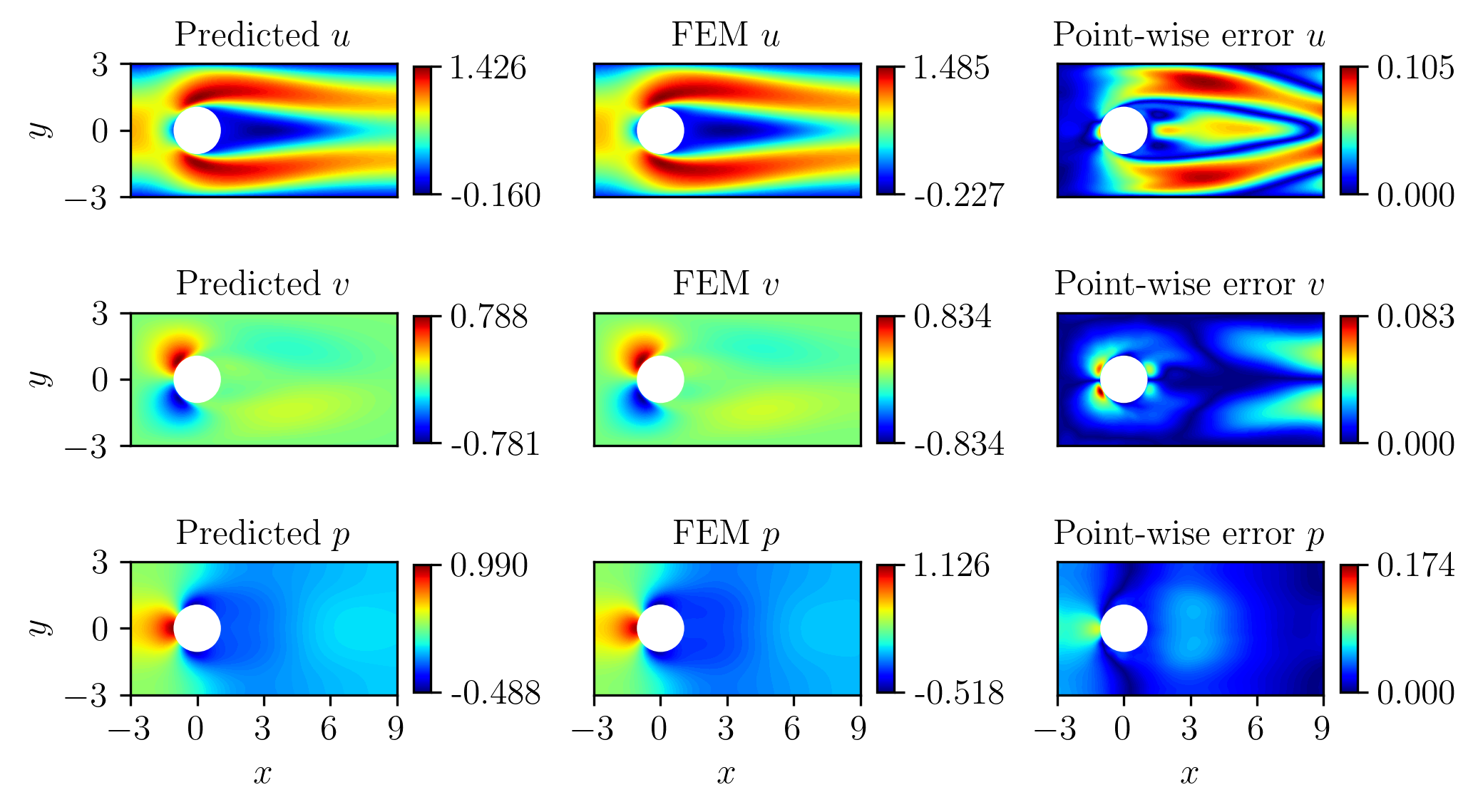}
    \caption{{\bf Solution of flow past a cylinder at Re = 50}: Solution predicted by FastVPINNs, FEM solution and point-wise errors for $u$, $v$ and $p$.}
    \label{fig:fpc_solution}
\end{figure*}

To demonstrate our framework's capability in handling complex geometries, we solved the flow past a cylinder problem using a mesh with skewed quadrilateral elements, as illustrated in Fig.~\ref{fig:fpc_mesh}. The computational domain comprises of 553 elements, each utilizing 9 test functions and 16 quadrature points, resulting in a total of 8,848 quadrature points across the domain. We sample 1,673 boundary points to impose the boundary conditions. For this geometry, we employed a neural network architecture consisting of 7 hidden layers with 20 neurons each. The network was trained for 250,000 steps using an initial learning rate of $2.9 \times 10^{-3}$, coupled with an exponential learning rate scheduler with a decay step of 2200 and a decay rate of 0.985) to ensure stable convergence. Fig.~\ref{fig:fpc_loss} shows the training progression, showing the evolution of total loss and its components over the training iterations. We set the weighting coefficients $\alpha$, $\beta$, and $\gamma$ in Eq.~\eqref{eq:variational_loss_NSE} as 10.
The results, presented in Fig.~\ref{fig:fpc_solution}, demonstrate good agreement between the FastVPINNs predictions and the FEM solution for both velocity and pressure components. This validates our framework's ability to accurately capture complex flow patterns around the cylinder, including the wake region.
Table~\ref{tab:fpc_errors} quantifies the accuracy of our approach, presenting the  $L^2_{rel}$ errors for velocity and pressure components. These results underscore FastVPINNs' effectiveness in solving fluid dynamics problems on complex geometries, achieving good accuracy while maintaining computational efficiency.
\begin{table}[H]
\centering
\begin{tabular}{|c|c|}
\hline
\textbf{Component} & \textbf{Relative $L^2$ Error} \\
\hline
$u$ (x-velocity) & $5.60 \times 10^{-2}$ \\
\hline
$v$ (y-velocity) & $9.58 \times 10^{-2}$ \\
\hline
$p$ (pressure) & $1.20 \times 10^{-1}$ \\
\hline
\end{tabular}
\caption{Relative $L^2$ errors for FastVPINNs solution of flow past cylinder at $Re=50$.}
\label{tab:fpc_errors}
\end{table}

\section{Inverse Problems}
\begin{figure*}[tbp!]
    \centering
    \includegraphics[width=0.7\textwidth]{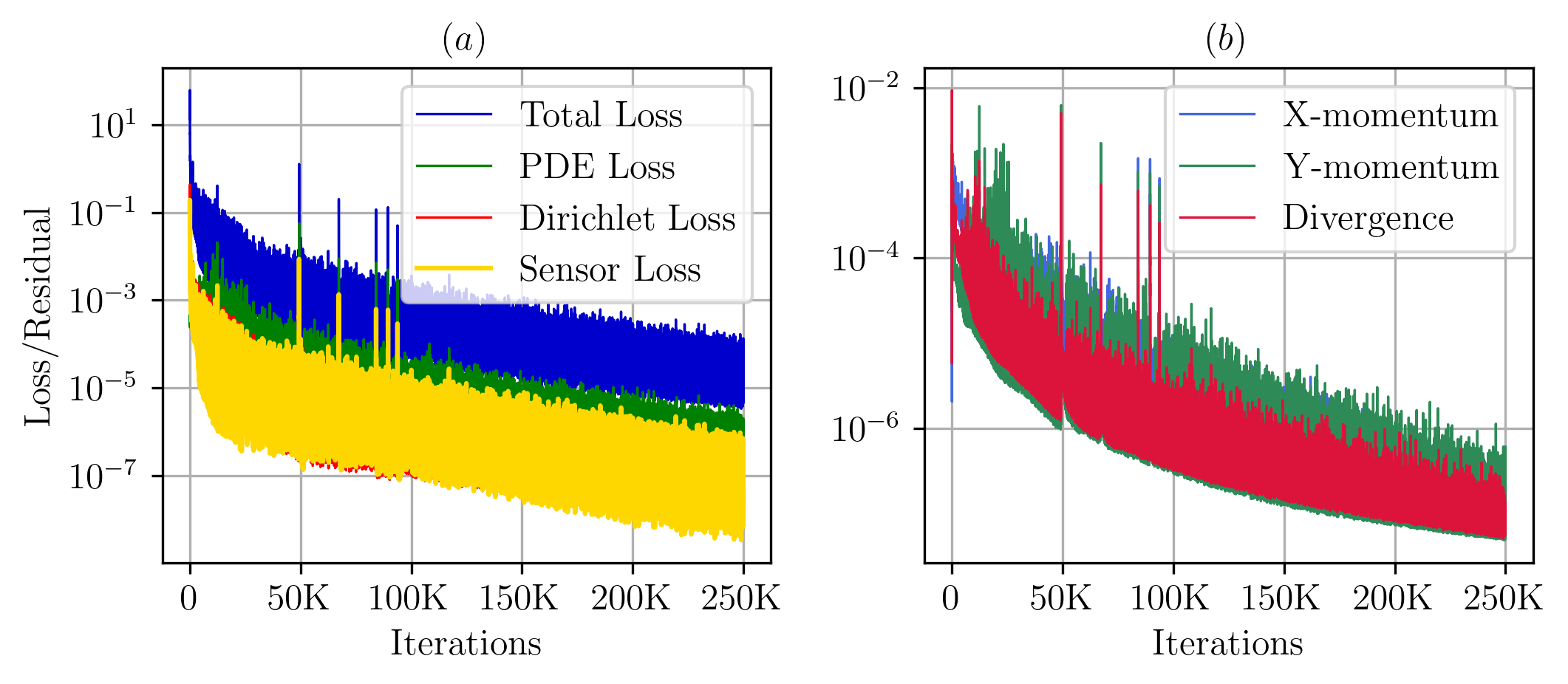}
    \caption{{\bf Inverse problem with flow past backward facing step}: (a) PDE loss, Dirichlet boundary loss, sensor loss and total loss during training, (b) PDE loss split into its components: x-momentum residual, y-momentum residual and divergence loss.}
    \label{fig:backwardstep_inverse_loss}
\end{figure*}

\begin{figure*}[ht!]
    \centering
    \includegraphics[width=0.7\textwidth]{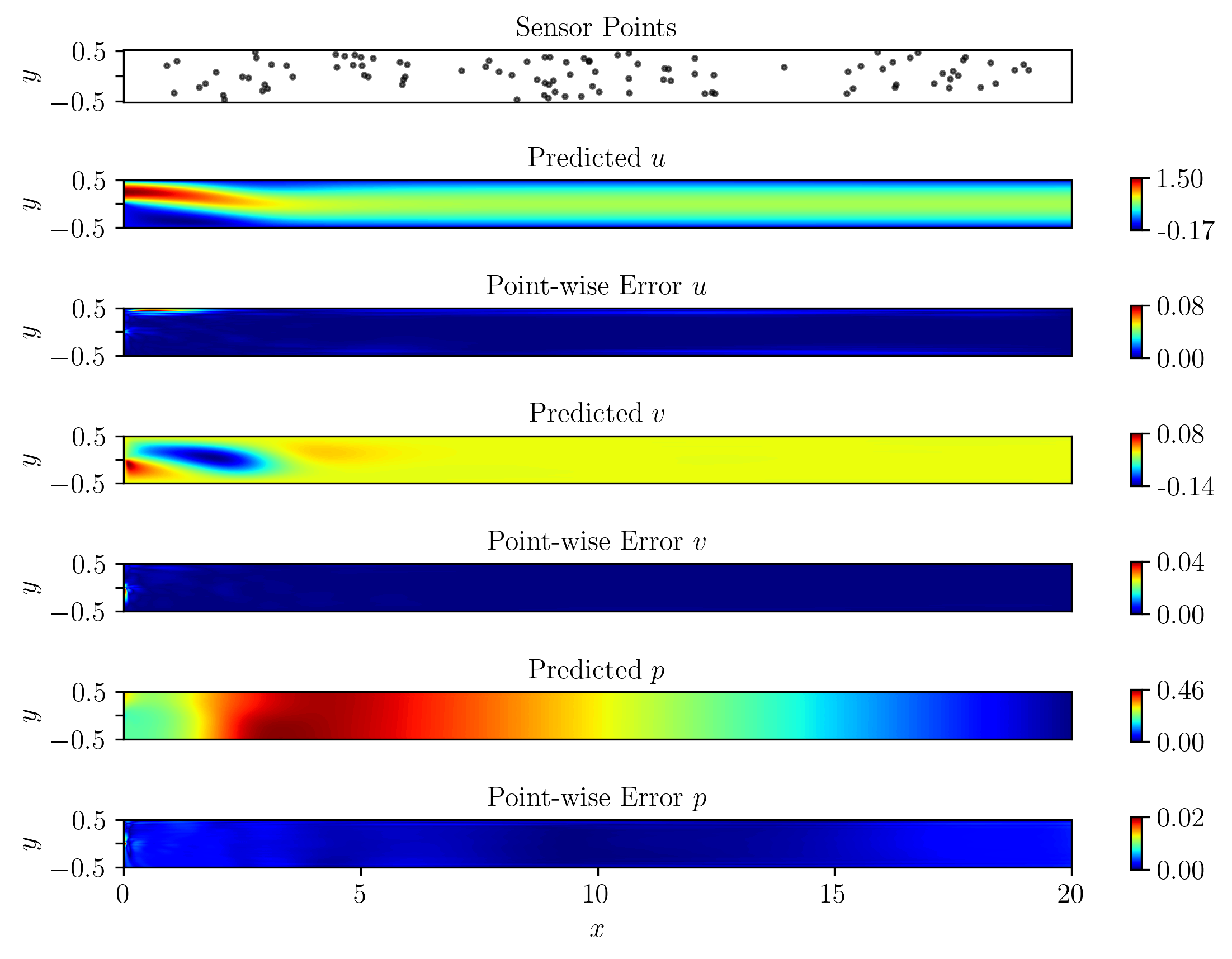}
    \caption{{\bf Inverse problem with flow past backward facing step}: Sensor points, predicted values and point-wise errors for $u$, $v$ and $p$.}
    \label{fig:backwardstep_inverse_solution}
\end{figure*}

To demonstrate the capability of FastVPINNs in solving inverse problems, we apply our framework to identify the Reynolds number in a flow through a backward-facing step. Inverse problems in fluid dynamics often involve estimating parameters of the governing PDEs, such as the Reynolds number in Navier-Stokes equations. In FastVPINNs, this is achieved by treating the target parameter as a trainable variable within the neural network architecture. Our computational domain spans $[0, 20]$ in the x-dimension and $[-0.5, 0.5]$ in the y-dimension, with an inlet velocity profile in the x-dimension given by,
\begin{equation*}
    u = 24y(0.5-y) \quad \text{for} \ y \in [0, 0.5]
\end{equation*}
along the left wall. The actual Reynolds number for this flow is 200, which our model aims to predict. We discretize the domain into 100 elements $(20 \times 5)$, using 25 test functions and 64 quadrature points per element, resulting in a total of 6400 quadrature points. Additionally, we sample 800 boundary points and set a boundary $\tau$ of 100. To predict the Reynolds number, we utilize velocity data from 100 randomly distributed sensor points within the domain. Our neural network architecture comprises 8 hidden layers with 50 neurons each, trained for 250,000 epochs using an initial learning rate of $1.9 \times 10^{-3}$, coupled with an exponential decay scheduler (decay step of 1500, decay rate of 0.985). Fig.~\ref{fig:backwardstep_inverse_loss}(a) and (b) illustrate the evolution of various loss components during the training process, providing insight into the convergence behavior of our model. The sensor loss, which represents the discrepancy between the neural network predictions and the data from sensor points, is shown in Fig.~\ref{fig:backwardstep_inverse_loss}(a). Fig.~\ref{fig:backwardstep_inverse_solution} presents the final predicted velocity and pressure fields, demonstrating the ability of the framework to capture flow patterns while simultaneously estimating the Reynolds number. Table~\ref{tab:inverse_errors} quantifies the accuracy of our predictions, showing the $L^2_{rel}$ errors for velocity and pressure components when the initial guess for the Reynolds number was 120. It took 1800s to train the model for 250,000 iterations.

\begin{figure*}[tbp!]
    \centering
    \includegraphics[width=0.6\textwidth]{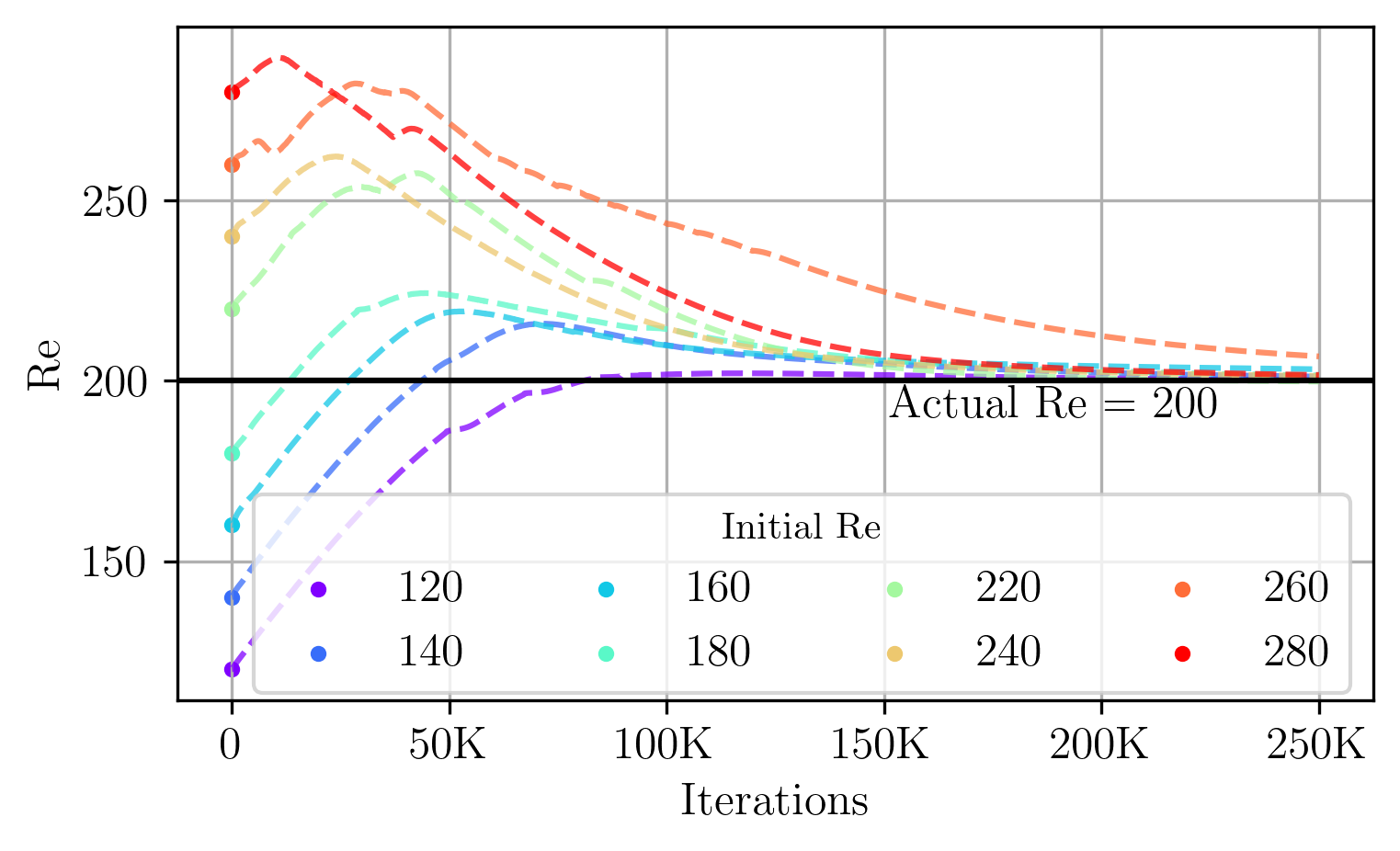}
    \caption{{\bf Inverse problem with flow past backward facing step}: Predicted Reynolds number during training for different initial guesses.}
    \label{fig:backwardstep_inverse_loss_Re}
\end{figure*}

To assess the robustness of our approach, we initialized the Reynolds number estimate with values ranging from 120 to 280. Fig.~\ref{fig:backwardstep_inverse_loss_Re} illustrates the convergence of these estimates to the true value of 200, highlighting the framework's ability to accurately determine the Reynolds number regardless of the initial guess. These results demonstrate the effectiveness of FastVPINNs in solving inverse problems in fluid dynamics, showing promise for applications in parameter estimation and flow characterization in complex geometries.
\begin{table}[ht!]
    \centering
    \begin{tabular}{|c|c|}
    \hline
    \textbf{Component} & \textbf{Relative $L^2$ Error} \\
    \hline
    $u$ (x-velocity) & $6.88 \times 10^{-3}$ \\
    \hline
    $v$ (y-velocity) & $3.16 \times 10^{-2}$ \\
    \hline
    $p$ (pressure) & $5.48 \times 10^{-3}$ \\
    \hline
    \end{tabular}
    \caption{Relative $L^2$ errors for the inverse problem solution in the backward-facing step flow.}
    \label{tab:inverse_errors}
\end{table}

\section{Conclusion}\label{sec:conclusion}

In this work, we have successfully extended the FastVPINNs framework to solve vector-valued problems, with a particular focus on the incompressible Navier-Stokes equations. We validated our approach through a comprehensive set of numerical experiments, including problems such as the Kovasznay flow, lid-driven cavity flow, and flow past a backward-facing step. Our implementation demonstrates significant improvements in both computational efficiency and accuracy compared to existing methods. Notably, in our comparisons with existing literature on PINNs for the Kovasznay flow and Falkner-Skan boundary layer problems, we achieved a 2x speedup in training time while maintaining comparable or improved accuracy relative to the PINNs code mentioned in the literature. The framework's ability to handle complex geometries was showcased by solving the flow past a cylinder problem on a domain composed of skewed quadrilateral elements. Furthermore, we demonstrated the versatility of FastVPINNs by successfully applying it to an inverse problem, accurately predicting the Reynolds number in a flow past a backward-facing step configuration. This capability highlights the potential of our method for parameter estimation and flow characterization tasks. The combination of improved computational efficiency, accuracy, and the ability to handle both forward and inverse problems positions FastVPINNs as a promising tool for a wide range of fluid dynamics applications, from simple geometries to complex, real-world scenarios.
This current work serves as a technology demonstrator and would be able to extend to other real-world engineering applications like turbulent flow in a windfarm or multiphase flows in pipelines.

\section{Acknowledgement}
We thank Shell India Markets Private Limited for providing funding for this project. The authors thank MHRD Grant No.STARS-1/388 (SPADE) for partial support.

\section*{Authors Declarations}

\subsection*{Conflict of Interest Declaration}
The authors report that this study received funding from Shell India Private Limited. Ankit Tyagi, Abhineet Gupta, and Suranjan Sarkar were employees of Shell India Private Limited during the study. These authors were involved in validation, providing resources, data curation, writing the original draft, and making the decision to submit for publication. The remaining authors declare no conflicts of interest.

\subsection*{Author Contributions}
\textbf{Thivin Anandh}: Conceptualization; Methodology; Software; Validation; Data curation; Investigation; Supervision, Writing - original draft.  Writing - review and editing; \textbf{Divij Ghose}: Conceptualization; Validation; Methodology; Data curation; Visualization; Writing - original draft;  Writing - review and editing; \textbf{Ankit Tyagi}: Validation; Data curation; Writing - original draft. \textbf{Abhineet Gupta}: Validation; Data curation; Writing - original draft. \textbf{Suranjan Sarkar}: Resources; Validation; Writing - review and editing; Supervision;\textbf{Sashikumaar Ganesan}: Conceptualization; Formal analysis; Validation; Resources; Writing - review and editing; Supervision; Project administration; Funding acquisition.

\subsection*{Data Availability}
The data that supports the findings of this study are openly available at \url{https://github.com/cmgcds/fastvpinns}(FastVPINNs GitHub repository).

\appendix
\section{}
\subsection{Burgers' equation}\label{sec:appx_burgers}
The steady-state, viscous Burgers' equation in the 2D domain $\Omega$ is given by,
\begin{align}
    \begin{split}
    u \frac{\partial u}{\partial x} + v \frac{\partial u}{\partial y} - \nu \left( \frac{\partial^2 u}{\partial x^2} + \frac{\partial^2 u}{\partial y^2} \right) &= f_x \quad \text{in} \ \Omega,\\
    u \frac{\partial v}{\partial x} + v \frac{\partial v}{\partial y} - \nu \left( \frac{\partial^2 v}{\partial x^2} + \frac{\partial^2 v}{\partial y^2} \right) &= f_y \quad \text{in} \ \Omega.        
    \end{split}
    \label{eqn:burgers2d}
\end{align}
Now, the variational form of~\eqref{eqn:burgers2d} can be obtained by multiplying the equation with a vector valued test function $\mathbf{\phi} = [\phi_1 , \phi_2]^T$ and integrating it over the domain
\begin{align}
    \int_{\Omega} \left( u \frac{\partial u}{\partial x} + v \frac{\partial u}{\partial y} \right) \phi_x \, d\Omega \nonumber \\
    + \int_{\Omega} \nu \left( \frac{\partial u}{\partial x}\frac{\partial \phi_x}{\partial x}
    + \frac{\partial u}{\partial y} \frac{\partial \phi_x}{\partial y}\right) \, d\Omega 
    &= \int_{\Omega} f_{x}\phi_x \, d\Omega, \\
    \int_{\Omega} \left( u \frac{\partial v}{\partial x} + v \frac{\partial v}{\partial y} \right) \phi_y \, d\Omega \nonumber \\
    + \int_{\Omega} \nu \left( \frac{\partial v}{\partial x}\frac{\partial \phi_y}{\partial x}
    + \frac{\partial v}{\partial y} \frac{\partial \phi_y}{\partial y}\right) \, d\Omega 
    &= \int_{\Omega} f_{y}\phi_y \, d\Omega,
\end{align}
\subsection{Kovasznay flow}\label{sec:appx_kovasznay}
The accuracy and training time comparison between FastVPINNs and NSFNets, for a network configuration of 4 hidden layers with 50 neurons each, is given in the table below.

\renewcommand{\arraystretch}{1.2}
\begin{table}[ht!]
\centering
\begin{tabular}{|c|c|c|}
\hline
Metric & FastVPINNs & NSFnets \\ \hline
\textbf{$L^2_{rel} (u)$} & $(4.90 \pm 3.60) \times 10^{-3}$ & $(7.20 \pm 1.00) \times 10^{-3}$ \\ \hline
\textbf{$L^2_{rel} (v)$} & $(2.38 \pm 1.63) \times 10^{-2}$ & $(5.84 \pm 0.81) \times 10^{-2}$ \\ \hline
\textbf{$L^2_{rel} (p)$} & $(6.30 \pm 1.70) \times 10^{-3}$ & $(2.71 \pm 0.41) \times 10^{-2}$ \\ \hline
\begin{tabular}[c]{@{}c@{}}Training Time for\\ 1K Iterations (s)\end{tabular} & \textbf{5.55} & 13.60 \\ \hline
\end{tabular}
\caption{Comparison of accuracy and training time between FastVPINNs and NSFnets for Kovasznay flow at $\text{Re}=40$.}
\label{tab:kovasznay_accuracy_50x4}
\end{table}

\subsection{Pseudocode for Loss computation of Incompressible Navier-Stokes Equation}\label{sec:appx_nse2d}

\newpage
\noindent\rule{0.5\textwidth}{1pt}  
Algorithm 1: FastVPINNs loss for Incompressible Navier-Stokes Equation
\noindent\rule{0.5\textwidth}{1pt}  
\begin{lstlisting}[language=Python, label={lst:algorithm_3}]

# Define acronyms:
# TM: tf.linalg.matvec (Tensor-Matrix multiplication)
# TO: tf.transpose(tf.linalg.matvec) (Transposed Output of Tensor-Matrix multiplication)
# test_v: test_velocity
# test_v_x: test_velocity_grad_x
# test_v_y: test_velocity_grad_y
# test_p: test_pressure

def train_step():
    # Obtain solution for the quadrature points in entire domain
    sol = model(quad_points_in_domain)
    u_nn, v_nn, p_nn = sol[:, 0], sol[:, 1], sol[:, 2]
    
    # Obtain gradients
    u_x, u_y = model.get_gradients(u_nn).reshape(Ne, Nq)
    v_x, v_y = model.get_gradients(v_nn).reshape(Ne, Nq)
    
    # X-momentum equation components
    diffusion_x = (1.0 / Re) * (TO(test_v_x, u_x) 
                    + TO(test_v_y, u_y))
    conv_x      = TO(test_v, u_x * u_nn) 
                    + TO(test_v, u_y * v_nn)
    pressure_x  = TO(test_v_x, p_nn)
    
    # Y-momentum equation components
    diffusion_y = (1.0 / Re) * (TO(test_v_x, v_x) 
                    + TO(test_v_y, v_y))
    conv_y      = TO(test_v, v_x * u_nn) 
                    + TO(test_v, v_y * v_nn)
    pressure_y  = TO(test_v_y, p_nn)
    
    # Continuity equation
    divergence  = TO(test_p, u_x) 
                    + TO(test_p, v_y)
    
    # Compute residuals
    residual_x  = diffusion_x + conv_x - pressure_x
    residual_y  = diffusion_y + conv_y - pressure_y
    
    # Compute mean squared residuals
    residual_x  = reduce_mean(square(residual_x), axis=0)
    residual_y  = reduce_mean(square(residual_y), axis=0)
    divergence  = reduce_mean(square(divergence), axis=0)
    
    # Compute total residual with penalties
    residual_cells = (residual_u_penalty * residual_x 
                    + residual_v_penalty * residual_y 
                    + divergence_penalty * divergence)
    
    # Compute variational loss
    variational_loss = reduce_sum(residual_cells)
    
    # Compute boundary losses (Dirichlet and Neumann)
    dirichlet_loss = compute_dirichlet_loss(u_nn, v_nn, g_1, g_2)
    neumann_loss   = compute_neumann_loss(u_x, u_y, v_x, v_y, h_1, h_2)
    
    # Compute total loss
    total_loss = (variational_loss 
                + tau_D * dirichlet_loss 
                + tau_N * neumann_loss)

    return total_loss

    

\end{lstlisting}
\noindent\rule{0.5\textwidth}{1pt} \label{app1}

\section*{REFERENCES}
\nocite{*}
\bibliography{aipsamp}

\begin{thebibliography}{35}%
\makeatletter
\providecommand \@ifxundefined [1]{%
 \@ifx{#1\undefined}
}%
\providecommand \@ifnum [1]{%
 \ifnum #1\expandafter \@firstoftwo
 \else \expandafter \@secondoftwo
 \fi
}%
\providecommand \@ifx [1]{%
 \ifx #1\expandafter \@firstoftwo
 \else \expandafter \@secondoftwo
 \fi
}%
\providecommand \natexlab [1]{#1}%
\providecommand \enquote  [1]{``#1''}%
\providecommand \bibnamefont  [1]{#1}%
\providecommand \bibfnamefont [1]{#1}%
\providecommand \citenamefont [1]{#1}%
\providecommand \href@noop [0]{\@secondoftwo}%
\providecommand \href [0]{\begingroup \@sanitize@url \@href}%
\providecommand \@href[1]{\@@startlink{#1}\@@href}%
\providecommand \@@href[1]{\endgroup#1\@@endlink}%
\providecommand \@sanitize@url [0]{\catcode `\\12\catcode `\$12\catcode `\&12\catcode `\#12\catcode `\^12\catcode `\_12\catcode `\%12\relax}%
\providecommand \@@startlink[1]{}%
\providecommand \@@endlink[0]{}%
\providecommand \url  [0]{\begingroup\@sanitize@url \@url }%
\providecommand \@url [1]{\endgroup\@href {#1}{\urlprefix }}%
\providecommand \urlprefix  [0]{URL }%
\providecommand \Eprint [0]{\href }%
\providecommand \doibase [0]{http://dx.doi.org/}%
\providecommand \selectlanguage [0]{\@gobble}%
\providecommand \bibinfo  [0]{\@secondoftwo}%
\providecommand \bibfield  [0]{\@secondoftwo}%
\providecommand \translation [1]{[#1]}%
\providecommand \BibitemOpen [0]{}%
\providecommand \bibitemStop [0]{}%
\providecommand \bibitemNoStop [0]{.\EOS\space}%
\providecommand \EOS [0]{\spacefactor3000\relax}%
\providecommand \BibitemShut  [1]{\csname bibitem#1\endcsname}%
\let\auto@bib@innerbib\@empty
\bibitem [{\citenamefont {Raissi}, \citenamefont {Perdikaris},\ and\ \citenamefont {Karniadakis}(2019)}]{raissi2019physics}%
  \BibitemOpen
  \bibfield  {author} {\bibinfo {author} {\bibfnamefont {M.}~\bibnamefont {Raissi}}, \bibinfo {author} {\bibfnamefont {P.}~\bibnamefont {Perdikaris}}, \ and\ \bibinfo {author} {\bibfnamefont {G.~E.}\ \bibnamefont {Karniadakis}},\ }\bibfield  {title} {\enquote {\bibinfo {title} {Physics-informed neural networks: A deep learning framework for solving forward and inverse problems involving nonlinear partial differential equations},}\ }\href@noop {} {\bibfield  {journal} {\bibinfo  {journal} {Journal of Computational physics}\ }\textbf {\bibinfo {volume} {378}},\ \bibinfo {pages} {686--707} (\bibinfo {year} {2019})}\BibitemShut {NoStop}%
\bibitem [{\citenamefont {de~Wolff}\ \emph {et~al.}(2021)\citenamefont {de~Wolff}, \citenamefont {Carrillo}, \citenamefont {Mart{\'\i}},\ and\ \citenamefont {Sanchez-Pi}}]{de2021assessing}%
  \BibitemOpen
  \bibfield  {author} {\bibinfo {author} {\bibfnamefont {T.}~\bibnamefont {de~Wolff}}, \bibinfo {author} {\bibfnamefont {H.}~\bibnamefont {Carrillo}}, \bibinfo {author} {\bibfnamefont {L.}~\bibnamefont {Mart{\'\i}}}, \ and\ \bibinfo {author} {\bibfnamefont {N.}~\bibnamefont {Sanchez-Pi}},\ }\bibfield  {title} {\enquote {\bibinfo {title} {Assessing physics informed neural networks in ocean modelling and climate change applications},}\ }in\ \href@noop {} {\emph {\bibinfo {booktitle} {AI: Modeling Oceans and Climate Change Workshop at ICLR 2021}}}\ (\bibinfo {year} {2021})\BibitemShut {NoStop}%
\bibitem [{\citenamefont {Haghighat}\ \emph {et~al.}(2021)\citenamefont {Haghighat}, \citenamefont {Raissi}, \citenamefont {Moure}, \citenamefont {Gomez},\ and\ \citenamefont {Juanes}}]{haghighat2021physics}%
  \BibitemOpen
  \bibfield  {author} {\bibinfo {author} {\bibfnamefont {E.}~\bibnamefont {Haghighat}}, \bibinfo {author} {\bibfnamefont {M.}~\bibnamefont {Raissi}}, \bibinfo {author} {\bibfnamefont {A.}~\bibnamefont {Moure}}, \bibinfo {author} {\bibfnamefont {H.}~\bibnamefont {Gomez}}, \ and\ \bibinfo {author} {\bibfnamefont {R.}~\bibnamefont {Juanes}},\ }\bibfield  {title} {\enquote {\bibinfo {title} {A physics-informed deep learning framework for inversion and surrogate modeling in solid mechanics},}\ }\href@noop {} {\bibfield  {journal} {\bibinfo  {journal} {Computer Methods in Applied Mechanics and Engineering}\ }\textbf {\bibinfo {volume} {379}},\ \bibinfo {pages} {113741} (\bibinfo {year} {2021})}\BibitemShut {NoStop}%
\bibitem [{\citenamefont {Karniadakis}\ \emph {et~al.}(2021)\citenamefont {Karniadakis}, \citenamefont {Kevrekidis}, \citenamefont {Lu}, \citenamefont {Perdikaris}, \citenamefont {Wang},\ and\ \citenamefont {Yang}}]{karniadakis2021physics}%
  \BibitemOpen
  \bibfield  {author} {\bibinfo {author} {\bibfnamefont {G.~E.}\ \bibnamefont {Karniadakis}}, \bibinfo {author} {\bibfnamefont {I.~G.}\ \bibnamefont {Kevrekidis}}, \bibinfo {author} {\bibfnamefont {L.}~\bibnamefont {Lu}}, \bibinfo {author} {\bibfnamefont {P.}~\bibnamefont {Perdikaris}}, \bibinfo {author} {\bibfnamefont {S.}~\bibnamefont {Wang}}, \ and\ \bibinfo {author} {\bibfnamefont {L.}~\bibnamefont {Yang}},\ }\bibfield  {title} {\enquote {\bibinfo {title} {Physics-informed machine learning},}\ }\href@noop {} {\bibfield  {journal} {\bibinfo  {journal} {Nature Reviews Physics}\ }\textbf {\bibinfo {volume} {3}},\ \bibinfo {pages} {422--440} (\bibinfo {year} {2021})}\BibitemShut {NoStop}%
\bibitem [{\citenamefont {Cai}\ \emph {et~al.}(2021)\citenamefont {Cai}, \citenamefont {Mao}, \citenamefont {Wang}, \citenamefont {Yin},\ and\ \citenamefont {Karniadakis}}]{cai2021physics}%
  \BibitemOpen
  \bibfield  {author} {\bibinfo {author} {\bibfnamefont {S.}~\bibnamefont {Cai}}, \bibinfo {author} {\bibfnamefont {Z.}~\bibnamefont {Mao}}, \bibinfo {author} {\bibfnamefont {Z.}~\bibnamefont {Wang}}, \bibinfo {author} {\bibfnamefont {M.}~\bibnamefont {Yin}}, \ and\ \bibinfo {author} {\bibfnamefont {G.~E.}\ \bibnamefont {Karniadakis}},\ }\bibfield  {title} {\enquote {\bibinfo {title} {Physics-informed neural networks (pinns) for fluid mechanics: A review},}\ }\href@noop {} {\bibfield  {journal} {\bibinfo  {journal} {Acta Mechanica Sinica}\ }\textbf {\bibinfo {volume} {37}},\ \bibinfo {pages} {1727--1738} (\bibinfo {year} {2021})}\BibitemShut {NoStop}%
\bibitem [{\citenamefont {Arthurs}\ and\ \citenamefont {King}(2021)}]{arthurs2021active}%
  \BibitemOpen
  \bibfield  {author} {\bibinfo {author} {\bibfnamefont {C.~J.}\ \bibnamefont {Arthurs}}\ and\ \bibinfo {author} {\bibfnamefont {A.~P.}\ \bibnamefont {King}},\ }\bibfield  {title} {\enquote {\bibinfo {title} {Active training of physics-informed neural networks to aggregate and interpolate parametric solutions to the navier-stokes equations},}\ }\href@noop {} {\bibfield  {journal} {\bibinfo  {journal} {Journal of Computational Physics}\ }\textbf {\bibinfo {volume} {438}},\ \bibinfo {pages} {110364} (\bibinfo {year} {2021})}\BibitemShut {NoStop}%
\bibitem [{\citenamefont {Raissi}, \citenamefont {Yazdani},\ and\ \citenamefont {Karniadakis}(2020)}]{raissi2020hidden}%
  \BibitemOpen
  \bibfield  {author} {\bibinfo {author} {\bibfnamefont {M.}~\bibnamefont {Raissi}}, \bibinfo {author} {\bibfnamefont {A.}~\bibnamefont {Yazdani}}, \ and\ \bibinfo {author} {\bibfnamefont {G.~E.}\ \bibnamefont {Karniadakis}},\ }\bibfield  {title} {\enquote {\bibinfo {title} {Hidden fluid mechanics: Learning velocity and pressure fields from flow visualizations},}\ }\href@noop {} {\bibfield  {journal} {\bibinfo  {journal} {Science}\ }\textbf {\bibinfo {volume} {367}},\ \bibinfo {pages} {1026--1030} (\bibinfo {year} {2020})}\BibitemShut {NoStop}%
\bibitem [{\citenamefont {Mahmoudabadbozchelou}, \citenamefont {Karniadakis},\ and\ \citenamefont {Jamali}(2022)}]{mahmoudabadbozchelou2022nn}%
  \BibitemOpen
  \bibfield  {author} {\bibinfo {author} {\bibfnamefont {M.}~\bibnamefont {Mahmoudabadbozchelou}}, \bibinfo {author} {\bibfnamefont {G.~E.}\ \bibnamefont {Karniadakis}}, \ and\ \bibinfo {author} {\bibfnamefont {S.}~\bibnamefont {Jamali}},\ }\bibfield  {title} {\enquote {\bibinfo {title} {nn-pinns: Non-newtonian physics-informed neural networks for complex fluid modeling},}\ }\href@noop {} {\bibfield  {journal} {\bibinfo  {journal} {Soft Matter}\ }\textbf {\bibinfo {volume} {18}},\ \bibinfo {pages} {172--185} (\bibinfo {year} {2022})}\BibitemShut {NoStop}%
\bibitem [{\citenamefont {Biswas}\ and\ \citenamefont {Anand}(2023)}]{biswas2023three}%
  \BibitemOpen
  \bibfield  {author} {\bibinfo {author} {\bibfnamefont {S.~K.}\ \bibnamefont {Biswas}}\ and\ \bibinfo {author} {\bibfnamefont {N.}~\bibnamefont {Anand}},\ }\bibfield  {title} {\enquote {\bibinfo {title} {Three-dimensional laminar flow using physics informed deep neural networks},}\ }\href@noop {} {\bibfield  {journal} {\bibinfo  {journal} {Physics of Fluids}\ }\textbf {\bibinfo {volume} {35}} (\bibinfo {year} {2023})}\BibitemShut {NoStop}%
\bibitem [{\citenamefont {Jin}\ \emph {et~al.}(2021)\citenamefont {Jin}, \citenamefont {Cai}, \citenamefont {Li},\ and\ \citenamefont {Karniadakis}}]{jin2021nsfnets}%
  \BibitemOpen
  \bibfield  {author} {\bibinfo {author} {\bibfnamefont {X.}~\bibnamefont {Jin}}, \bibinfo {author} {\bibfnamefont {S.}~\bibnamefont {Cai}}, \bibinfo {author} {\bibfnamefont {H.}~\bibnamefont {Li}}, \ and\ \bibinfo {author} {\bibfnamefont {G.~E.}\ \bibnamefont {Karniadakis}},\ }\bibfield  {title} {\enquote {\bibinfo {title} {Nsfnets (navier-stokes flow nets): Physics-informed neural networks for the incompressible navier-stokes equations},}\ }\href@noop {} {\bibfield  {journal} {\bibinfo  {journal} {Journal of Computational Physics}\ }\textbf {\bibinfo {volume} {426}},\ \bibinfo {pages} {109951} (\bibinfo {year} {2021})}\BibitemShut {NoStop}%
\bibitem [{\citenamefont {Eivazi}\ \emph {et~al.}(2022)\citenamefont {Eivazi}, \citenamefont {Tahani}, \citenamefont {Schlatter},\ and\ \citenamefont {Vinuesa}}]{eivazi2022physics}%
  \BibitemOpen
  \bibfield  {author} {\bibinfo {author} {\bibfnamefont {H.}~\bibnamefont {Eivazi}}, \bibinfo {author} {\bibfnamefont {M.}~\bibnamefont {Tahani}}, \bibinfo {author} {\bibfnamefont {P.}~\bibnamefont {Schlatter}}, \ and\ \bibinfo {author} {\bibfnamefont {R.}~\bibnamefont {Vinuesa}},\ }\bibfield  {title} {\enquote {\bibinfo {title} {Physics-informed neural networks for solving reynolds-averaged navier--stokes equations},}\ }\href@noop {} {\bibfield  {journal} {\bibinfo  {journal} {Physics of Fluids}\ }\textbf {\bibinfo {volume} {34}} (\bibinfo {year} {2022})}\BibitemShut {NoStop}%
\bibitem [{\citenamefont {Kag}, \citenamefont {Seshasayanan},\ and\ \citenamefont {Gopinath}(2022)}]{kag2022physics}%
  \BibitemOpen
  \bibfield  {author} {\bibinfo {author} {\bibfnamefont {V.}~\bibnamefont {Kag}}, \bibinfo {author} {\bibfnamefont {K.}~\bibnamefont {Seshasayanan}}, \ and\ \bibinfo {author} {\bibfnamefont {V.}~\bibnamefont {Gopinath}},\ }\bibfield  {title} {\enquote {\bibinfo {title} {Physics-informed data based neural networks for two-dimensional turbulence},}\ }\href@noop {} {\bibfield  {journal} {\bibinfo  {journal} {Physics of Fluids}\ }\textbf {\bibinfo {volume} {34}} (\bibinfo {year} {2022})}\BibitemShut {NoStop}%
\bibitem [{\citenamefont {Li}\ and\ \citenamefont {Feng}(2022)}]{li2022dynamic}%
  \BibitemOpen
  \bibfield  {author} {\bibinfo {author} {\bibfnamefont {S.}~\bibnamefont {Li}}\ and\ \bibinfo {author} {\bibfnamefont {X.}~\bibnamefont {Feng}},\ }\bibfield  {title} {\enquote {\bibinfo {title} {Dynamic weight strategy of physics-informed neural networks for the 2d navier--stokes equations},}\ }\href@noop {} {\bibfield  {journal} {\bibinfo  {journal} {Entropy}\ }\textbf {\bibinfo {volume} {24}},\ \bibinfo {pages} {1254} (\bibinfo {year} {2022})}\BibitemShut {NoStop}%
\bibitem [{\citenamefont {Zhang}, \citenamefont {Braga-Neto},\ and\ \citenamefont {Gildin}(2024)}]{zhang2024physics}%
  \BibitemOpen
  \bibfield  {author} {\bibinfo {author} {\bibfnamefont {J.}~\bibnamefont {Zhang}}, \bibinfo {author} {\bibfnamefont {U.}~\bibnamefont {Braga-Neto}}, \ and\ \bibinfo {author} {\bibfnamefont {E.}~\bibnamefont {Gildin}},\ }\bibfield  {title} {\enquote {\bibinfo {title} {Physics-informed neural networks for multiphase flow in porous media considering dual shocks and interphase solubility},}\ }\href@noop {} {\bibfield  {journal} {\bibinfo  {journal} {Energy \& Fuels}\ } (\bibinfo {year} {2024})}\BibitemShut {NoStop}%
\bibitem [{\citenamefont {Wassing}, \citenamefont {Langer},\ and\ \citenamefont {Bekemeyer}(2024)}]{wassing2024physics}%
  \BibitemOpen
  \bibfield  {author} {\bibinfo {author} {\bibfnamefont {S.}~\bibnamefont {Wassing}}, \bibinfo {author} {\bibfnamefont {S.}~\bibnamefont {Langer}}, \ and\ \bibinfo {author} {\bibfnamefont {P.}~\bibnamefont {Bekemeyer}},\ }\bibfield  {title} {\enquote {\bibinfo {title} {Physics-informed neural networks for parametric compressible euler equations},}\ }\href@noop {} {\bibfield  {journal} {\bibinfo  {journal} {Computers \& Fluids}\ }\textbf {\bibinfo {volume} {270}},\ \bibinfo {pages} {106164} (\bibinfo {year} {2024})}\BibitemShut {NoStop}%
\bibitem [{\citenamefont {Gu}\ \emph {et~al.}(2024)\citenamefont {Gu}, \citenamefont {Qin}, \citenamefont {Xu},\ and\ \citenamefont {Chen}}]{gu2024physics}%
  \BibitemOpen
  \bibfield  {author} {\bibinfo {author} {\bibfnamefont {L.}~\bibnamefont {Gu}}, \bibinfo {author} {\bibfnamefont {S.}~\bibnamefont {Qin}}, \bibinfo {author} {\bibfnamefont {L.}~\bibnamefont {Xu}}, \ and\ \bibinfo {author} {\bibfnamefont {R.}~\bibnamefont {Chen}},\ }\bibfield  {title} {\enquote {\bibinfo {title} {Physics-informed neural networks with domain decomposition for the incompressible navier--stokes equations},}\ }\href@noop {} {\bibfield  {journal} {\bibinfo  {journal} {Physics of Fluids}\ }\textbf {\bibinfo {volume} {36}} (\bibinfo {year} {2024})}\BibitemShut {NoStop}%
\bibitem [{\citenamefont {Kharazmi}, \citenamefont {Zhang},\ and\ \citenamefont {Karniadakis}(2019)}]{kharazmi2019variational}%
  \BibitemOpen
  \bibfield  {author} {\bibinfo {author} {\bibfnamefont {E.}~\bibnamefont {Kharazmi}}, \bibinfo {author} {\bibfnamefont {Z.}~\bibnamefont {Zhang}}, \ and\ \bibinfo {author} {\bibfnamefont {G.~E.}\ \bibnamefont {Karniadakis}},\ }\bibfield  {title} {\enquote {\bibinfo {title} {Variational physics-informed neural networks for solving partial differential equations},}\ }\href@noop {} {\bibfield  {journal} {\bibinfo  {journal} {arXiv preprint arXiv:1912.00873}\ } (\bibinfo {year} {2019})}\BibitemShut {NoStop}%
\bibitem [{\citenamefont {Khodayi-Mehr}\ and\ \citenamefont {Zavlanos}(2020)}]{khodayi2020varnet}%
  \BibitemOpen
  \bibfield  {author} {\bibinfo {author} {\bibfnamefont {R.}~\bibnamefont {Khodayi-Mehr}}\ and\ \bibinfo {author} {\bibfnamefont {M.}~\bibnamefont {Zavlanos}},\ }\bibfield  {title} {\enquote {\bibinfo {title} {Varnet: Variational neural networks for the solution of partial differential equations},}\ }in\ \href@noop {} {\emph {\bibinfo {booktitle} {Learning for Dynamics and Control}}}\ (\bibinfo {organization} {PMLR},\ \bibinfo {year} {2020})\ pp.\ \bibinfo {pages} {298--307}\BibitemShut {NoStop}%
\bibitem [{\citenamefont {Kharazmi}, \citenamefont {Zhang},\ and\ \citenamefont {Karniadakis}(2021)}]{kharazmi2021hp}%
  \BibitemOpen
  \bibfield  {author} {\bibinfo {author} {\bibfnamefont {E.}~\bibnamefont {Kharazmi}}, \bibinfo {author} {\bibfnamefont {Z.}~\bibnamefont {Zhang}}, \ and\ \bibinfo {author} {\bibfnamefont {G.~E.}\ \bibnamefont {Karniadakis}},\ }\bibfield  {title} {\enquote {\bibinfo {title} {hp-vpinns: Variational physics-informed neural networks with domain decomposition},}\ }\href@noop {} {\bibfield  {journal} {\bibinfo  {journal} {Computer Methods in Applied Mechanics and Engineering}\ }\textbf {\bibinfo {volume} {374}},\ \bibinfo {pages} {113547} (\bibinfo {year} {2021})}\BibitemShut {NoStop}%
\bibitem [{\citenamefont {Anandh}\ \emph {et~al.}(2024)\citenamefont {Anandh}, \citenamefont {Ghose}, \citenamefont {Jain},\ and\ \citenamefont {Ganesan}}]{anandh2024fastvpinns}%
  \BibitemOpen
  \bibfield  {author} {\bibinfo {author} {\bibfnamefont {T.}~\bibnamefont {Anandh}}, \bibinfo {author} {\bibfnamefont {D.}~\bibnamefont {Ghose}}, \bibinfo {author} {\bibfnamefont {H.}~\bibnamefont {Jain}}, \ and\ \bibinfo {author} {\bibfnamefont {S.}~\bibnamefont {Ganesan}},\ }\bibfield  {title} {\enquote {\bibinfo {title} {{FastVPINNs: Tensor-Driven Acceleration of VPINNs for Complex Geometries}},}\ }\href@noop {} {\bibfield  {journal} {\bibinfo  {journal} {arXiv preprint arXiv:2404.12063}\ } (\bibinfo {year} {2024})}\BibitemShut {NoStop}%
\bibitem [{\citenamefont {Anandh}, \citenamefont {Ghose},\ and\ \citenamefont {Ganesan}(2024)}]{anandh2024fastvpinnsjoss}%
  \BibitemOpen
  \bibfield  {author} {\bibinfo {author} {\bibfnamefont {T.}~\bibnamefont {Anandh}}, \bibinfo {author} {\bibfnamefont {D.}~\bibnamefont {Ghose}}, \ and\ \bibinfo {author} {\bibfnamefont {S.}~\bibnamefont {Ganesan}},\ }\bibfield  {title} {\enquote {\bibinfo {title} {{FastVPINNs: An efficient tensor-based Python library for solving partial differential equations using hp-Variational Physics Informed Neural Networks}},}\ }\href@noop {} {\bibfield  {journal} {\bibinfo  {journal} {Journal of Open Source Software}\ }\textbf {\bibinfo {volume} {9}},\ \bibinfo {pages} {6764} (\bibinfo {year} {2024})}\BibitemShut {NoStop}%
\bibitem [{\citenamefont {Ghose}, \citenamefont {Anandh},\ and\ \citenamefont {Ganesan}(2024)}]{ghose2024fastvpinns}%
  \BibitemOpen
  \bibfield  {author} {\bibinfo {author} {\bibfnamefont {D.}~\bibnamefont {Ghose}}, \bibinfo {author} {\bibfnamefont {T.}~\bibnamefont {Anandh}}, \ and\ \bibinfo {author} {\bibfnamefont {S.}~\bibnamefont {Ganesan}},\ }\bibfield  {title} {\enquote {\bibinfo {title} {Fastvpinns: A fast, versatile and robust variational pinns framework for forward and inverse problems in science},}\ }in\ \href@noop {} {\emph {\bibinfo {booktitle} {ICLR 2024 Workshop on AI4DifferentialEquations In Science}}}\ (\bibinfo {year} {2024})\BibitemShut {NoStop}%
\bibitem [{\citenamefont {Kharazmi}(2023)}]{hp_vpinns_github}%
  \BibitemOpen
  \bibfield  {author} {\bibinfo {author} {\bibfnamefont {E.}~\bibnamefont {Kharazmi}},\ }\href {https://github.com/ehsankharazmi/hp-VPINNs} {\enquote {\bibinfo {title} {Github: {hp-VPINNs}: hp-variational physics-informed neural networks},}\ } (\bibinfo {year} {2023})\BibitemShut {NoStop}%
\bibitem [{\citenamefont {Abadi}\ \emph {et~al.}(2015)\citenamefont {Abadi}, \citenamefont {Agarwal}, \citenamefont {Barham}, \citenamefont {Brevdo}, \citenamefont {Chen}, \citenamefont {Citro}, \citenamefont {Corrado}, \citenamefont {Davis}, \citenamefont {Dean}, \citenamefont {Devin}, \citenamefont {Ghemawat}, \citenamefont {Goodfellow}, \citenamefont {Harp}, \citenamefont {Irving}, \citenamefont {Isard}, \citenamefont {Jia}, \citenamefont {Jozefowicz}, \citenamefont {Kaiser}, \citenamefont {Kudlur}, \citenamefont {Levenberg}, \citenamefont {Man\'{e}}, \citenamefont {Monga}, \citenamefont {Moore}, \citenamefont {Murray}, \citenamefont {Olah}, \citenamefont {Schuster}, \citenamefont {Shlens}, \citenamefont {Steiner}, \citenamefont {Sutskever}, \citenamefont {Talwar}, \citenamefont {Tucker}, \citenamefont {Vanhoucke}, \citenamefont {Vasudevan}, \citenamefont {Vi\'{e}gas}, \citenamefont {Vinyals}, \citenamefont {Warden}, \citenamefont {Wattenberg}, \citenamefont {Wicke}, \citenamefont {Yu},\ and\ \citenamefont
  {Zheng}}]{tensorflow2015-whitepaper}%
  \BibitemOpen
  \bibfield  {author} {\bibinfo {author} {\bibfnamefont {M.}~\bibnamefont {Abadi}}, \bibinfo {author} {\bibfnamefont {A.}~\bibnamefont {Agarwal}}, \bibinfo {author} {\bibfnamefont {P.}~\bibnamefont {Barham}}, \bibinfo {author} {\bibfnamefont {E.}~\bibnamefont {Brevdo}}, \bibinfo {author} {\bibfnamefont {Z.}~\bibnamefont {Chen}}, \bibinfo {author} {\bibfnamefont {C.}~\bibnamefont {Citro}}, \bibinfo {author} {\bibfnamefont {G.~S.}\ \bibnamefont {Corrado}}, \bibinfo {author} {\bibfnamefont {A.}~\bibnamefont {Davis}}, \bibinfo {author} {\bibfnamefont {J.}~\bibnamefont {Dean}}, \bibinfo {author} {\bibfnamefont {M.}~\bibnamefont {Devin}}, \bibinfo {author} {\bibfnamefont {S.}~\bibnamefont {Ghemawat}}, \bibinfo {author} {\bibfnamefont {I.}~\bibnamefont {Goodfellow}}, \bibinfo {author} {\bibfnamefont {A.}~\bibnamefont {Harp}}, \bibinfo {author} {\bibfnamefont {G.}~\bibnamefont {Irving}}, \bibinfo {author} {\bibfnamefont {M.}~\bibnamefont {Isard}}, \bibinfo {author} {\bibfnamefont {Y.}~\bibnamefont {Jia}}, \bibinfo
  {author} {\bibfnamefont {R.}~\bibnamefont {Jozefowicz}}, \bibinfo {author} {\bibfnamefont {L.}~\bibnamefont {Kaiser}}, \bibinfo {author} {\bibfnamefont {M.}~\bibnamefont {Kudlur}}, \bibinfo {author} {\bibfnamefont {J.}~\bibnamefont {Levenberg}}, \bibinfo {author} {\bibfnamefont {D.}~\bibnamefont {Man\'{e}}}, \bibinfo {author} {\bibfnamefont {R.}~\bibnamefont {Monga}}, \bibinfo {author} {\bibfnamefont {S.}~\bibnamefont {Moore}}, \bibinfo {author} {\bibfnamefont {D.}~\bibnamefont {Murray}}, \bibinfo {author} {\bibfnamefont {C.}~\bibnamefont {Olah}}, \bibinfo {author} {\bibfnamefont {M.}~\bibnamefont {Schuster}}, \bibinfo {author} {\bibfnamefont {J.}~\bibnamefont {Shlens}}, \bibinfo {author} {\bibfnamefont {B.}~\bibnamefont {Steiner}}, \bibinfo {author} {\bibfnamefont {I.}~\bibnamefont {Sutskever}}, \bibinfo {author} {\bibfnamefont {K.}~\bibnamefont {Talwar}}, \bibinfo {author} {\bibfnamefont {P.}~\bibnamefont {Tucker}}, \bibinfo {author} {\bibfnamefont {V.}~\bibnamefont {Vanhoucke}}, \bibinfo {author}
  {\bibfnamefont {V.}~\bibnamefont {Vasudevan}}, \bibinfo {author} {\bibfnamefont {F.}~\bibnamefont {Vi\'{e}gas}}, \bibinfo {author} {\bibfnamefont {O.}~\bibnamefont {Vinyals}}, \bibinfo {author} {\bibfnamefont {P.}~\bibnamefont {Warden}}, \bibinfo {author} {\bibfnamefont {M.}~\bibnamefont {Wattenberg}}, \bibinfo {author} {\bibfnamefont {M.}~\bibnamefont {Wicke}}, \bibinfo {author} {\bibfnamefont {Y.}~\bibnamefont {Yu}}, \ and\ \bibinfo {author} {\bibfnamefont {X.}~\bibnamefont {Zheng}},\ }\href {https://www.tensorflow.org/} {\enquote {\bibinfo {title} {{TensorFlow}: Large-scale machine learning on heterogeneous systems},}\ } (\bibinfo {year} {2015}),\ \bibinfo {note} {software available from tensorflow.org}\BibitemShut {NoStop}%
\bibitem [{\citenamefont {Babu{\v{s}}ka}(1973)}]{babuvska1973finite}%
  \BibitemOpen
  \bibfield  {author} {\bibinfo {author} {\bibfnamefont {I.}~\bibnamefont {Babu{\v{s}}ka}},\ }\bibfield  {title} {\enquote {\bibinfo {title} {The finite element method with lagrangian multipliers},}\ }\href@noop {} {\bibfield  {journal} {\bibinfo  {journal} {Numerische Mathematik}\ }\textbf {\bibinfo {volume} {20}},\ \bibinfo {pages} {179--192} (\bibinfo {year} {1973})}\BibitemShut {NoStop}%
\bibitem [{\citenamefont {Brezzi}(1974)}]{brezzi1974existence}%
  \BibitemOpen
  \bibfield  {author} {\bibinfo {author} {\bibfnamefont {F.}~\bibnamefont {Brezzi}},\ }\bibfield  {title} {\enquote {\bibinfo {title} {On the existence, uniqueness and approximation of saddle-point problems arising from lagrangian multipliers},}\ }\href@noop {} {\bibfield  {journal} {\bibinfo  {journal} {Publications des s{\'e}minaires de math{\'e}matiques et informatique de Rennes}\ ,\ \bibinfo {pages} {1--26}} (\bibinfo {year} {1974})}\BibitemShut {NoStop}%
\bibitem [{\citenamefont {Ganesan}\ and\ \citenamefont {Tobiska}(2017)}]{ganesan2017finite}%
  \BibitemOpen
  \bibfield  {author} {\bibinfo {author} {\bibfnamefont {S.}~\bibnamefont {Ganesan}}\ and\ \bibinfo {author} {\bibfnamefont {L.}~\bibnamefont {Tobiska}},\ }\href@noop {} {\emph {\bibinfo {title} {Finite elements: Theory and algorithms}}}\ (\bibinfo  {publisher} {Cambridge University Press},\ \bibinfo {year} {2017})\BibitemShut {NoStop}%
\bibitem [{\citenamefont {Kingma}(2014)}]{kingma2014adam}%
  \BibitemOpen
  \bibfield  {author} {\bibinfo {author} {\bibfnamefont {D.~P.}\ \bibnamefont {Kingma}},\ }\bibfield  {title} {\enquote {\bibinfo {title} {Adam: A method for stochastic optimization},}\ }\href@noop {} {\bibfield  {journal} {\bibinfo  {journal} {arXiv preprint arXiv:1412.6980}\ } (\bibinfo {year} {2014})}\BibitemShut {NoStop}%
\bibitem [{\citenamefont {Kovasznay}(1948)}]{Kovasznay_1948}%
  \BibitemOpen
  \bibfield  {author} {\bibinfo {author} {\bibfnamefont {L.~I.~G.}\ \bibnamefont {Kovasznay}},\ }\bibfield  {title} {\enquote {\bibinfo {title} {Laminar flow behind a two-dimensional grid},}\ }\href {\doibase 10.1017/S0305004100023999} {\bibfield  {journal} {\bibinfo  {journal} {Mathematical Proceedings of the Cambridge Philosophical Society}\ }\textbf {\bibinfo {volume} {44}},\ \bibinfo {pages} {58–62} (\bibinfo {year} {1948})}\BibitemShut {NoStop}%
\bibitem [{\citenamefont {Wilbrandt}\ \emph {et~al.}(2017)\citenamefont {Wilbrandt}, \citenamefont {Bartsch}, \citenamefont {Ahmed}, \citenamefont {Alia}, \citenamefont {Anker}, \citenamefont {Blank}, \citenamefont {Caiazzo}, \citenamefont {Ganesan}, \citenamefont {Giere}, \citenamefont {Matthies} \emph {et~al.}}]{wilbrandt2017parmoon}%
  \BibitemOpen
  \bibfield  {author} {\bibinfo {author} {\bibfnamefont {U.}~\bibnamefont {Wilbrandt}}, \bibinfo {author} {\bibfnamefont {C.}~\bibnamefont {Bartsch}}, \bibinfo {author} {\bibfnamefont {N.}~\bibnamefont {Ahmed}}, \bibinfo {author} {\bibfnamefont {N.}~\bibnamefont {Alia}}, \bibinfo {author} {\bibfnamefont {F.}~\bibnamefont {Anker}}, \bibinfo {author} {\bibfnamefont {L.}~\bibnamefont {Blank}}, \bibinfo {author} {\bibfnamefont {A.}~\bibnamefont {Caiazzo}}, \bibinfo {author} {\bibfnamefont {S.}~\bibnamefont {Ganesan}}, \bibinfo {author} {\bibfnamefont {S.}~\bibnamefont {Giere}}, \bibinfo {author} {\bibfnamefont {G.}~\bibnamefont {Matthies}},  \emph {et~al.},\ }\bibfield  {title} {\enquote {\bibinfo {title} {Parmoon—a modernized program package based on mapped finite elements},}\ }\href@noop {} {\bibfield  {journal} {\bibinfo  {journal} {Computers \& Mathematics with Applications}\ }\textbf {\bibinfo {volume} {74}},\ \bibinfo {pages} {74--88} (\bibinfo {year} {2017})}\BibitemShut {NoStop}%
\bibitem [{\citenamefont {Gartling}(1990)}]{gartling1990test}%
  \BibitemOpen
  \bibfield  {author} {\bibinfo {author} {\bibfnamefont {D.~K.}\ \bibnamefont {Gartling}},\ }\bibfield  {title} {\enquote {\bibinfo {title} {A test problem for outflow boundary conditions—flow over a backward-facing step},}\ }\href@noop {} {\bibfield  {journal} {\bibinfo  {journal} {International Journal for Numerical Methods in Fluids}\ }\textbf {\bibinfo {volume} {11}},\ \bibinfo {pages} {953--967} (\bibinfo {year} {1990})}\BibitemShut {NoStop}%
\bibitem [{\citenamefont {Falkneb}\ and\ \citenamefont {Skan}(1931)}]{falkneb1931lxxxv}%
  \BibitemOpen
  \bibfield  {author} {\bibinfo {author} {\bibfnamefont {V.}~\bibnamefont {Falkneb}}\ and\ \bibinfo {author} {\bibfnamefont {S.~W.}\ \bibnamefont {Skan}},\ }\bibfield  {title} {\enquote {\bibinfo {title} {Lxxxv. solutions of the boundary-layer equations},}\ }\href@noop {} {\bibfield  {journal} {\bibinfo  {journal} {The London, Edinburgh, and Dublin Philosophical Magazine and Journal of Science}\ }\textbf {\bibinfo {volume} {12}},\ \bibinfo {pages} {865--896} (\bibinfo {year} {1931})}\BibitemShut {NoStop}%
\bibitem [{\citenamefont {KTH-FlowAI}(2023)}]{kth_flowai_pinns_2023}%
  \BibitemOpen
  \bibfield  {author} {\bibinfo {author} {\bibnamefont {KTH-FlowAI}},\ }\href {https://github.com/KTH-FlowAI/Physics-informed-neural-networks-for-solving-Reynolds-averaged-Navier-Stokes-equations} {\enquote {\bibinfo {title} {Physics-informed neural networks for solving reynolds-averaged navier-stokes equations},}\ } (\bibinfo {year} {2023})\BibitemShut {NoStop}%
\bibitem [{\citenamefont {Jalili}\ \emph {et~al.}(2024)\citenamefont {Jalili}, \citenamefont {Jang}, \citenamefont {Jadidi}, \citenamefont {Giustini}, \citenamefont {Keshmiri},\ and\ \citenamefont {Mahmoudi}}]{jalili2024physics}%
  \BibitemOpen
  \bibfield  {author} {\bibinfo {author} {\bibfnamefont {D.}~\bibnamefont {Jalili}}, \bibinfo {author} {\bibfnamefont {S.}~\bibnamefont {Jang}}, \bibinfo {author} {\bibfnamefont {M.}~\bibnamefont {Jadidi}}, \bibinfo {author} {\bibfnamefont {G.}~\bibnamefont {Giustini}}, \bibinfo {author} {\bibfnamefont {A.}~\bibnamefont {Keshmiri}}, \ and\ \bibinfo {author} {\bibfnamefont {Y.}~\bibnamefont {Mahmoudi}},\ }\bibfield  {title} {\enquote {\bibinfo {title} {Physics-informed neural networks for heat transfer prediction in two-phase flows},}\ }\href@noop {} {\bibfield  {journal} {\bibinfo  {journal} {International Journal of Heat and Mass Transfer}\ }\textbf {\bibinfo {volume} {221}},\ \bibinfo {pages} {125089} (\bibinfo {year} {2024})}\BibitemShut {NoStop}%
\bibitem [{\citenamefont {Bararnia}\ and\ \citenamefont {Esmaeilpour}(2022)}]{bararnia2022application}%
  \BibitemOpen
  \bibfield  {author} {\bibinfo {author} {\bibfnamefont {H.}~\bibnamefont {Bararnia}}\ and\ \bibinfo {author} {\bibfnamefont {M.}~\bibnamefont {Esmaeilpour}},\ }\bibfield  {title} {\enquote {\bibinfo {title} {On the application of physics informed neural networks (pinn) to solve boundary layer thermal-fluid problems},}\ }\href@noop {} {\bibfield  {journal} {\bibinfo  {journal} {International Communications in Heat and Mass Transfer}\ }\textbf {\bibinfo {volume} {132}},\ \bibinfo {pages} {105890} (\bibinfo {year} {2022})}\BibitemShut {NoStop}%
\end{thebibliography}%

\end{document}